\documentclass[conference]{IEEEtran}
\IEEEoverridecommandlockouts
\usepackage{cite}
\usepackage{amsmath,amssymb,amsfonts}
\usepackage{algorithmic}
\usepackage{graphicx}
\usepackage{textcomp}
\usepackage{xcolor}
\usepackage{comment}
\usepackage{siunitx}
\usepackage{lipsum}
\usepackage{multirow}

\newcommand\blfootnote[1]{%
  \begingroup
  \renewcommand\thefootnote{}\footnote{#1}%
  \addtocounter{footnote}{-1}%
  \endgroup
}
\def\BibTeX{{\rm B\kern-.05em{\sc i\kern-.025em b}\kern-.08em
    T\kern-.1667em\lower.7ex\hbox{E}\kern-.125emX}}

\usepackage{fancyhdr}
\begin{document}

\title{Estimating Personal Model Parameters from Utterances in Model-based Reminiscence\\
 % so delete this line
% \thanks{Identify applicable funding agency here. If none, delete this.}
}

\if0
\author{\IEEEauthorblockN{Shoki Sakai}
%\IEEEauthorblockA{\textit{Department of Informatics} \\
\IEEEauthorblockA{\textit{Department of Informatics,} \\ \textit{Graduated School of Integrated} \\ 
\textit{Science and Technology} \\
\textit{Shizuoka University}\\
Shizuoka, Japan \\
sakai.shoki.18@shizuoka.ac.jp}
\and
\IEEEauthorblockN{Kazuki Itabashi}
%\IEEEauthorblockA{\textit{dept. name of organization (of Aff.)} \\
\IEEEauthorblockA{\textit{Department of Informatics,} \\ \textit{Graduated School of Integrated} \\ 
\textit{Science and Technology} \\
\textit{Shizuoka University}\\
Shizuoka, Japan \\
itabashi.kazuki.15@shizuoka.ac.jp}
\and
\IEEEauthorblockN{Junya Morita}
%\IEEEauthorblockA{\textit{dept. name of organization (of Aff.)} \\
\IEEEauthorblockA{\textit{Department of Informatics,} \\ \textit{Graduated School of Integrated} \\ 
\textit{Science and Technology} \\
\textit{Shizuoka University}\\
Shizuoka, Japan \\
j-morita@inf.shizuoka.ac.jp}
}
\fi
\author{
    \IEEEauthorblockN{Shoki Sakai, Kazuki Itabashi, Junya Morita}
\IEEEauthorblockA{\textit{Department of Informatics, Graduated School of Integrated Science and Technology, Shizuoka University}\\
3-5-1 Johoku, Naka-ku, Hamamatsu, Japan \\
sakai.shoki.18@shizuoka.ac.jp, itabashi.kazuki.15@shizuoka.ac.jp,  j-morita@inf.shizuoka.ac.jp}
}

\maketitle
\thispagestyle{fancy}

%%%%%%%%%%%%%%%%%%%%%%%%%%%%%%%%%%%%%%%%%%%%%%%%%%%%%%%%%%%%
\begin{abstract}
%Reminiscence therapy is one method of mental health care that uses memory recollection accompanied with emotion.
%In this study, we aim to estimate model parameters that represent the user's internal state from utterances generated when using a reminiscence therapy support system.
%We explored methods for estimating the parameters necessary for simulating memory recollection based on arousal and valence by focusing on the content and prosody of the user's utterances. 
%As a result, the subjective rating of valence and the impression evaluation of a photograph were estimated from the user's prosodic information.
%This suggests the possibility of mapping user utterances to model parameters.
%記憶の回想を利用したメンタルヘルスケアの手法として回想法がある．本研究では回想法支援システム利用時に生成される発話からユーザの内部状態を表すモデルパラメータを推定することを目標とする．ユーザの発話の内容，韻律に注目し，覚醒度や感情に応じた記憶の遷移をシミュレートする際に必要なパラメータを推定する方法の探索を行った．その結果，韻律情報から気分の主観評定及び写真の印象評価が推定され，モデルパラメータと対応付けられる可能性が示唆された．

 %Recalling pleasant memories of one's past is one way to achieve well-being.
Reminiscence therapy is mental health care based on the recollection of memories.
%, and in a typical setting, participants recall their own memories from their old photos and discuss about them with each other.
However, the effectiveness of this method varies amongst individuals. 
To solve this problem, it is necessary to provide more personalized support; therefore, this study utilized a computational model of personal memory recollection based on a cognitive architecture adaptive control of thought-rational (ACT-R).
An ACT-R memory model reflecting the state of users is expected to facilitate personal recollection.
In this study, we proposed a method for estimating the internal states of users through repeated interactions with the memory model.
The model, which contains the lifelog of the user, presents a memory item (stimulus) to the user, and receives the response of the user to the stimulus, based on which it adjusts the internal parameters of the model.
Through the repetition of these processes, the parameters of the model will reflect the internal states of the user.
To confirm the feasibility of the proposed method, we analyzed utterances of users when using a system that incorporates this model.
The results confirmed the ability of the method to estimate the memory retrieval parameters of the model from the utterances of the user.
In addition, the ability of the method to estimate changes in the mood of the user caused by using the system was confirmed.
These results support the feasibility of the interactive method for estimating human internal states, which will eventually contribute to the ability to induce memory and emotions recall for our well-being.
%自身の過去の幸福な思い出を回想することはWell-beingにつながる一つの方法である．過去の記憶を思い出すことを利用したメンタルヘルスケアの手法として回想法がある．回想法は過去の写真などから自身の記憶を思い出し，語り合う手法である．回想法の問題点として，その効果に個人差が大きいということが指摘されている．この問題を解決するためにはより個人に合わせた回想支援が必要である．
%そこで本研究では回想法の中でも特に個人の記憶の回想に焦点を当てる．個人の回想を支援することはユーザの認知的な内部状態を捉えることで可能になるだろう．しかし，人間の内部状態を直接捉えることはできないので，現在のユーザの状態を推定する必要がある．
%本研究では認知モデルがユーザとのインタラクションを繰り返していく中でユーザの内部状態を推定する手法を提案する．
%認知モデルはユーザに対して何らかの刺激を提示し，それに対するユーザの反応を受け取る．モデルはユーザの反応からモデルの内部パラメータを調整する．このプロセスを繰り返すことでモデルの内部状態はユーザの認知的な内部状態を反映する．
%提案手法の実現性を確認するため，認知モデルを組み込んだシステム利用時のユーザ発話を分析した．その結果，ユーザの発話から認知モデルの内部パラメータの推定とユーザのシステム利用時の気分変化の推定の可能性が確認された．このことから本研究で提案したインタラクティブな人間の内部状態の推定の手法の実現性が支持されたと考えられる．
\end{abstract}

\blfootnote{This work was supported by JSPS KAKENHI Grant Numbers 20H05560 and 22H04861.} 
%%%%%%%%%%%%%%%%%%%%%%%%%%%%%%%%%%%%%%%%%%%%%%%%%%%%%%%%%%%%
\section{Introduction}
\label{s1}

Happiness or health is one of our universal and constant desires. 
According to the WHO Charter, ``health is a state of complete physical, mental, and social well-being and not merely the absence of disease or infirmity \cite{10665-268688}.''
This definition suggests the importance of being mentally and socially, as well as physically fit.
One of the ways to improve one's mental well-being is by  recollecting pleasant memories from the past.
%This feeling of recalling nostalgic memories of the past is called nostalgia.
The emotion that arises from recalling pleasant memories of the past is known as nostalgia across various cultures \cite{hepper2014pancultural}, and has been reported to principally be a positive emotion, although it also includes bittersweet feelings \cite{sedikides2004nostalgia}.

% 
% WHO憲章には「Health is a state of complete physical, mental and social well-being and not merely the absence of disease or infirmity.」とある．これは肉体だけでなく精神的，社会的にも充実していることが重要であるということを示している．精神的に充実する方法の一つとして過去の幸福な思い出を思い出すことが挙げられる．このように過去の幸福な思い出を想起することはnostalgiaと呼ばれる．nostalgiaはほろ苦い感情も含まれるが主にポジティブな感情であることが報告されている．また，nostalgiaはHepperらによって文化を超えて共通する概念であることが明らかにされている．\cite{hepper2014pancultural}

The effect of nostalgia has been applied to mental health care.
%Reminiscence therapy is one of the mental health care methods to recollect past memories and improve mental condition.
For example, Butler proposed the use of psychotherapies, in which people discuss their old experiences and memories from past photographs, music, and other media related to them, for mental health care \cite{butler1963life}.
Such memory-based therapies include life review and reminiscence. The former aims to reevaluate one's life by reflecting on the past, whereas the latter aims to achieve psychological stability by evoking nostalgic memories.
Reminiscence therapy has been demonstrated to promote positive emotions and suppress negative emotions for all age groups \cite{hallford2021remembering, bryant2005using} not only to improve cognitive function in the elderly and patients with dementia \cite{akhoondzadeh2014effect, gonzalez2015reminiscence}.
%過去の思い出を想起し，精神状態の向上を図るメンタルヘルスケアの手法の一つとして回想法がある．回想法はButlerによって提案された手法で，自身に関する過去の写真や音楽などから過去の経験や思い出を語り合う心理療法である\cite{butler1963life}．回想法には過去を振り返り自身の人生の再評価を狙うライフレビューと，懐かしい思い出の喚起による心理的な安定を狙うレミニッセンスがある．回想は高齢者や認知症患者への認知機能改善などの効果があることが報告されている\cite{akhoondzadeh2014effect, gonzalez2015reminiscence}．
%他にも，回想法は高齢者だけではなく，若い人たちには心理的資源やwell-beingの向上の効果があることも報告されている\cite{hallford2021remembering, Bryant2005}.．

%Memory recall is a primary activity in reminiscence therapy.
%Such mental health care will become more effective by modeling the internal state of the individual.
Among several types of reminiscence therapies, this study mainly focused on the use of personal recollections, rather than group recollection \cite{otake2009coimagination}, to resolve large individual differences in emotional factors involved in memory contents.
In addition, we assumed two axes of core affect (arousal and valence) \cite{russell2003core} as the internal factors affecting the memory  recollection of an individual \cite{bradley1992remembering}.
This assumption suggests that encouraging memory recollection based on the state of these internal variables should provide appropriate support for the reminiscence of an individual.
%回想法にはいくつかの種類があるが，本研究ではその中でも特に個人の回想に焦点を当てる．回想法の問題点としてその効果について個人差が大きいことが指摘されている．個人に合わせた支援を行うことで，回想法はより効果を発揮するようになるだろう．そのためには,個人ごとの記憶をモデル化する必要がある．記憶の回想に関与すると考えられる内部変数としては，生体内部の覚醒度や感情価を想定できる\cite{bradley1992remembering, nobata2005effects}．適切な回想の支援を行うためにはこれらの内部変数の状態に応じた回想の促しが必要であると考えられる．

%Cognitive architecture can be used to model such cognitive processes.
%In this study, we propose an interaction design that reproduces the internal state of a human during memory recollection based on the interaction between a human and a cognitive model.
%In order to confirm the feasibility of the proposed design, we analyzed utterance data during the use of a reminiscence therapy support system with a cognitive model.
To achieve this personalized support, this study utilized a computational model that represents the recollection process of personal memory. There is considerable research on the simulation of human cognitive processes using cognitive architectures \cite{mathews1998cognitive, fum2007cognitive}, which %\cite{anderson2009can,laird2017standard} 
is a modeling approach that employs a standardized structure to represent an individual process occurring during a specific task.
%Cognitive models are computer models that represent human cognitive processes.
In addition, these structures can interact with humans by providing a stimulus to the user and receiving a response to it.
Accordingly, the model reflects the internal state of the user by adjusting the parameters of the cognitive model through repeated interactions.

We believe that building a cognitive model that simulates the memory recollection of the user can provide support for personalized reminiscence.
To date, the authors have explored the method of tracing the user's internal state during memory recall.
Based on these previous studies, this study proposed an interaction design for a cognitive model that can trace the human internal state.
To confirm the feasibility of the proposed design, we analyzed the utterance data obtained when a reminiscence support system equipped with the cognitive model was used.
%人間の認知プロセスをモデル化するアプローチの一つとして認知アーキテクチャがある．認知アーキテクチャとは認知モデルを構築するフレームワークである．認知モデルは人間の認知プロセスを表現する計算機モデルである．認知モデルを用いて人間の内部状態を表現する研究は盛んに行われている.認知モデルはユーザに何らかの刺激を提示し，それに対する反応を受け取ることで人間とインタラクションを行う．このインタラクションを繰り返していくことでモデル内部のパラメータを調整していくことで，モデルはユーザの認知状態を反映したものとなる．ユーザの記憶の回想をシミュレートする認知モデルを構築することで，個人に合わせた回想の支援が実現できると考える．そこで本研究では，記憶回想時の人間の内部状態をシミュレートとすることを考える．そのために認知モデルが人間の内部状態を推定するためのインタラクションデザインを提案する．そして，提案デザインの実現性を確かめるため，認知モデルを搭載した回想法支援システムの利用時の発話データを分析する.

%%%%%%%%%%%%%%%%%%%%%%%%%%%%%%%%%%%%%%%%%%%%%%%%%%%%%%%%%%%%
\section{Related Works}
\label{s2}
%\section{関連研究}

This section reviews previous research relating to the memory recollection support method investigated in this study.
Section \ref{sub2a} describes the framework of the cognitive architecture.
Section \ref{sub2b} introduces research on systems that support individual memory recollections as the precedents of this study. Following this review, Section \ref{sub2c} clarifies the limitations of the previous studies, which necessitated the goal of this study.
%本研究では，回想の支援のために個人のモデル化を行う．そこで認知アーキテクチャを用いたモデル化を行う．\ref{sub2a}では本研究で用いる認知アーキテクチャとモデルについて説明する．次に認知アーキテクチャと認知モデルを用いた個人を支援するシステムの先行研究について\ref{sub2b}で紹介する．

%%%%%%%%%%%%%%%%%%%%%%%%%%%%%%%%%%%%%%%%%%%%%%%%%%%%%%%%%%%%
\subsection{Cognitive Architecture and Memory Models}
\label{sub2a}
%\subsection{認知アーキテクチャと記憶のモデル}
%In this study, we use a cognitive architecture to model cognitive processes for individuals.
%本研究では，人間個人の認知プロセスをモデル化するために，認知アーキテクチャを用いる．
As mentioned previously, cognitive architecture is a structure that represents general human cognitive processes.
This structure can simulate diverse cognitive functions for various tasks and individuals by constructing cognitive models on a common foundation. 
Among the various cognitive architectures (see \cite{Kotseruba} as a review), adaptive control of thought-rational (ACT-R) has been employed in various studies \cite{Anderson:2007}.
%This architecture is composed of an accumulation of results from psychological experiments.
The essential components of ACT-R have been implemented as a production system, and are composed of various modules for vision, motor movements, memory storage, and goal management.
The system can capture various aspects of human cognition by communicating information between the modules that correspond to each part of the human brain, indicating the ability of ACT-R to reproduce the human internal process.
%認知アーキテクチャは，人間の一般的な認知プロセスをモデル化するプラットフォームのことである．共通の基盤の上に，多様なタスクや個人に対応した人間の多様な認知機能をシミュレートすることが可能となる．様々な認知アーキテクチャがあるなかで，John. R. Andersonによって開発された認知アーキテクチャACT-Rは 多くの研究で用いられている\cite{anderson2009can}．ACT-Rは心理学実験の積み重ねによって構成されている．ACT-Rの基本的な部分はプロダクションシステムとして実装されている．また，視覚モジュールや運動モジュールなど様々なモジュールから構成されており，人間の各脳部位と対応付けられたモジュールが情報を伝達し合うことで人間の認知の多様な側面を捉えることができる．これらの特徴からACT-Rは人間の内部状態を再現するための有用な特徴を持つと考えられる．

Some studies have modeled individual humans using the ACT-R.
For example, Somer et al proposed the concept of the ``Cognitive Twin'' \cite{somerscognitive} and demonstrated the efficacy of ACT-R for personalized decision support by constructing and using a decision-making model of an individual.
%ACT-Rを用いた個人のモデルの構築も行われている．Somerらは，ACT-R を用いた Cognitive Twin という概念を提唱した\cite{somerscognitive}．彼らの研究では，個人の意思決定のモデルが作成され，それを利用することで，ACT-R によるパーソナライズされた意思決定支援の有効性が示された．
Such individualization of cognitive functions in the ACT-R was realized by the knowledge (memory) and the parameters involved in the use of knowledge. By manipulating these factors, the ACT-R model generates a variety of behaviors related to the same task.
The knowledge incorporated in the ACT-R consists of declarative and procedural knowledge. Declarative knowledge comprises units known as ``chunk'', and individual chunks are linked to form a semantic network through attributes.
Production rules, which involve procedural knowledge, are necessary for serially exploring such semantic networks.
Accordingly, the ACT-R model selects one of the attributes linked to the currently recalled knowledge, and induces its  to the next unit of knowledge.
%こういった ACT-R による認知機能の個別化は，モデルに搭載される知識（記憶）によって実現される．モデルに搭載する知識，あるいは知識の利用に関わるパラメータに応じて，ACT-R のモデルは同一の課題に対しても多様な振る舞いを生み出す．ACT-R の知識は宣言的知識と手続き知識から構成される．宣言的知識はチャンクと呼ばれる単位によって構成され，個々のチャンクが結びつくことで，属性を介した意味ネットワークが構成される．このようなネットワークを系列的に探索するためには，手続き知識であるプロダクションルールが必要になる．プロダクションルールは，現在想起されている知識と接続する属性から一つを選択することで，次の知識に遷移する．

This memory retrieval process can be diversified by manipulating the numerical parameters attached to the knowledge. Chunks and production rules have parameters called the activation and the utility, respectively.
The activations are affected by the frequency of memory recall and memory decay, whereas the utilities are affected by rewards and the learning rate.
When invoking declarative knowledge using a semantic network, the utility of the production rule is used in selecting attributes, and the activation determines the chunk to be retrieved when there are multiple available candidates among the selected attributes.
%上記のような ACT-R における記憶の検索プロセスにバリエーションをもたせるパラメータとして，チャンクに対する活性値，プロダクションルール対するユーティリティ値が用意されている．活性値は，記憶の利用頻度や記憶の減衰などの影響を受ける．ユーティリティ値は報酬値と学習率の影響を受ける．意味ネットワークからの宣言的記憶の呼び出しにおいては，属性の選択においてプロダクションルールのユーティリティが関与し，選択された属性のなかで複数の候補が合った際には活性値によって検索される知識が定まる．

%%%%%%%%%%%%%%%%%%%%%%%%%%%%%%%%%%%%%%%%%%%%%%%%%%%%%%%%%%%%
\subsection{Memory Support Systems Using Cognitive Models}
\label{sub2b}
%\subsection{認知モデルを用いたシステム}

This study follows a series of ACT-R based memory support systems developed by the authors' group. First, Morita et al \cite{morita2016modelbased} developed a photo slideshow system that supports reminiscence therapy and utilized the memory mechanism of the ACT-R described in Section \ref{sub2a}. This slide-show system includes a collection of private photos of users as a lifelog, and these photos are connected to each other via semantic links classified into four attributes: Who (person appearing in the photos), What (results of image recognition), When (timestamp), and Where (GPS information). 
The ACT-R model sequentially recollects photos by following these semantic links based on the utilities of each attribute and the activation of each memory item. 
The authors asserted that such memory transitions simulate mental time travel, which is vivid recalling of experiences in a time that is not the current time, as if experiencing time travel \cite{schacter2007remembering}.
Based on this assumption, they proposed a {\it model-based reminiscence} using visualizations generated by this mental time travel model.
%森田らは，上記のようなACT-Rの記憶の仕組みを活用した回想法支援のための写真スライドショーシステムを開発した\cite{森田純哉2015認知アーキテクチャを組み入れた写真スライドショーの開発}．また，現在の時間ではない時間の経験を，まるでタイムトラベルをする可能ように鮮明に思い浮かべることを意味するメンタルタイムトラベル\cite{schacter2007remembering}のモデル化を行い，構築したモデルによるモデルベース回想法を提案した\cite{morita2016modelbased}．

The model-based reminiscence involves the recollection of past memory based on dynamically changing photos in the form of a slideshow.
This method is considered to learn the cognitive and emotional states of the user, and present photos based on the learning (personalized) results.
%モデルベース回想法ではユーザはディスプレイ上にスライドショー形式で動的に切り替えられる写真によって回想を行う．このときモデルはユーザの認知的状態，情動的状態を学習し，学習結果に基づく刺激提示を行うと考えられる．
To utilize a user's reaction for such learning, Itabashi et al developed an interactive method based on an emotional model for monitoring the state of the user \cite{itabashi2020interactive}.
In this system, the authors assembled time-series data based on heart rate variability (HRV) and online subjective mood rating\footnote{The users rate their mood state for the currently displayed photo using the interface attached to the display of slideshow} as interactive parameters.
The feedback from the HRV (arousal) was directly mapped to a noise assigned to the activation of the photo, whereas the online subjective mood rating (valence) was mapped to the reward related to the rule for retrieving the photo.

ACT-R theoretically assumes that activation-based memory retrieval causes a concentration of retrieval on the most recent memory \cite{anderson1991reflections,lebiere2009balancing}. 
Therefore, the activation is considered to reflect the physiological arousal \cite{dancy2015using}.
By applying this assumption and utilizing HRV, the photo retrieval of the system proposed by Itabashi et al concentrates on a specific photo when the user is in a stress state (i.e., small HRV). Contrary, retrieval is not limited to recent photos but also older photos when arousal is low and a relaxed state is induced (i.e., high HRV).
%板橋は，森田らのシステムを改良するために，感情モデルに基づくユーザの状態の対話的なモニタリング手法を開発した\cite{itabashi2020interactive,itabashi2021master}．板橋のシステム\cite{itabashi2020interactive}では，対話的パラメータとして心拍変動の時系列データと，ユーザインタフェースによるフィードバックデータを設定した．ユーザは，図\ref{fig:interface}に示すインタフェースによって現在提示されている写真に対する気分状態を6段階で評定する．心拍変動とインタフェースによるフィードバックはそれぞれ，写真の活性値に付与されるノイズ値と，写真の検索ルールの報酬値と対応付けられている．活性値に基づく記憶の検索が行われることで，覚醒度が高い場合写真の撮影時期に基づいた直近の写真へ検索の集中が起こる．このような写真の検索は心拍間隔の短い緊張状態における記憶の回想と対応する．対して，覚醒度低い場合は直近の写真だけではなく古い写真への検索も行われる．このような写真の検索は，心拍が穏やかでリラックス状態における記憶の回想と対応する．

In addition, the interactive memory retrieval proposed by Itabashi et al has been applied in other contexts beyond a photo slideshow.
For example, Morita et al \cite{morita2022regulating} developed a web extension that mitigates negative emotions when using the Internet by building a model of memory and emotion using the ACT-R.
They focused on web advertising, including behavioral targeting, to naturally apply memory models in a web environment.
Their system displays regularly changing images on web pages, and these images affect the implicit memory process of the user.
They confirmed inhibition of negative memory recall during web use using a memory and emotion model, as well as physiological sensing to modulate memory retrieval.
%森田，ThanakitらはACT-Rを用いて記憶と感情のモデルを構築し，ウェブ利用時のネガティブな感情を和らげるためのウェブ拡張システムを開発した．彼らは，Web環境での記憶のモデルを自然に適用するため，行動ターゲティングなどのWeb広告に焦点を当てた．彼らのシステムはWebページに定期的に変更される画像を表示する．この画像はユーザの潜在記憶プロセスに影響を与える．彼らは記憶と感情のモデルと記憶検索を調節する生理的センシングを用いることで，ユーザがWeb利用時にネガティブな記憶の回想を抑制することを確認した．

%%%%%%%%%%%%%%%%%%%%%%%%%%%%%%%%%%%%%%%%%%%%%%%%%%%%%%%%%%%%
\subsection{Limitations of the Past Studies}\label{sub2c}
As discussed above, cognitive model-based personalized memory supports have been developed. However, the reliance of previous systems  on physiological data (i.e., HRV) that can only be measured using dedicated equipment has limited its application in various fields. 
In addition, the mood state rating used by Itabashi et al \cite{itabashi2020interactive} requires the user to intentionally operate the interface every time a picture is changed, which may inhibit natural reminiscence.
% 板橋らや森田らのシステムでは認知モデルへのフィードバックとして生理データである心拍変動や画面上に設置したユーザインタフェースを利用した．
%しかし，心拍変動の計測には専用の機材の用意が必要であることや，オンラインでのシステム利用という側面で問題がある．
%加えて，ユーザインタフェースによる気分状態の評価は写真が変わるごとにユーザに操作を要求するため回想を阻害する可能性がある．
To solve these limitations, this study proposed and tested an interaction design that indirectly estimates the internal states of the user by applying the machine-learning method to naturally produced user responses measured using a conventional device. 
%これらの課題を解決するため，我々は認知モデルへのフィードバックをユーザの音声によって行い，モデルがユーザとのインタラクションを繰り返すことでモデルのパラメータを調節する方法を提案する．
%モデルは \label{sub2a} で述べたように活性値と報酬値の2つのパラメータから記憶の検索を行う．
%モデルはこれらのパラメータをフィードバックを基に操作する．
%活性値はユーザの覚醒度に，報酬値はユーザの気分状態に対応している．
%そのため，音声からユーザの覚醒度と気分状態を推定する必要がある．
%%%%%%%%%%%%%%%%%%%%%%%%%%%%%%%%%%%%%%%%%%%%%%%%%%%%%%%%%%%%
\section{Interactive Estimation of User Model Parameters}
\label{s3}
%\section{ユーザモデルパラメータの対話的推定}
%To advance the interactive memory support approach presented in the previous section, this study proposes an interactive method that estimates human internal states, based on their response to the stimuli provided by the model.
%The proposed method of estimating human internal states, based on the human's responses to stimuli given to the human.

%本研究では，モデルと人間のインタラクションから人間の内部状態を推定する手法として，モデルが人間に与える刺激に対する反応を基に対話的な推定を行うことを提案する．提案するモデルと人間の関係図を図\ref{fig:image}に示す．

\begin{figure}[tb]
 \centering
 \includegraphics[keepaspectratio, width=50mm]
      {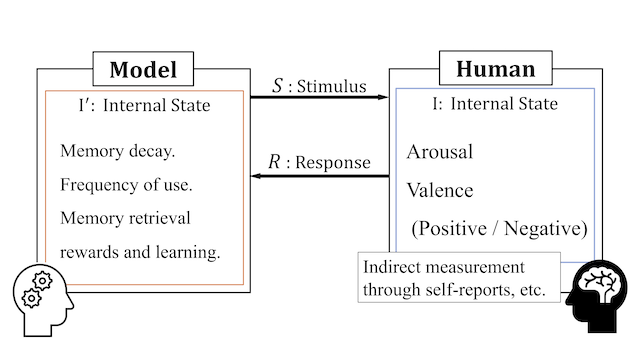}
 \caption{Scheme of the Human-Model interaction. The model provides a stimulus to a user. The user receives a stimulus and responds to it. The model updates its internal parameters based on the received response and determines a new stimulus.}
 \label{fig:image}
\end{figure}

The general framework of the human-model interaction adopted in this study is described in Fig.~\ref{fig:image}. In this framework, 
the model consists of several parameters for simulating memory recall as the internal states ($I'$). Based on the current $I'$, the model provides stimuli ($S$) to humans and receives their responses ($R$).
Subsequently, based on the human response, the model updates its internal state $I'$ and determines the next stimulus $S$ based on the updated $I'$.
By repeating this interaction, we expect that the internal state of the model ($I'$) will eventually reflect the human internal state ($I$).
%モデルは内部状態(I')として記憶の回想をシミュレートするためのパラメータを持つ．モデルは人間に刺激を与え ($S$)，それに対する人間の反応を受け取る ($R$)．人間の反応からモデルは自身の内部状態($I'$) を更新する．更新された$I'$を基に次の刺激を決定する．このやり取りを繰り返すことで$I'$ は人間の内部状態 ($I$) を反映するものになっていくと予想する．

\begin{figure}[tb]
 \centering
 \includegraphics[keepaspectratio, width=75mm]{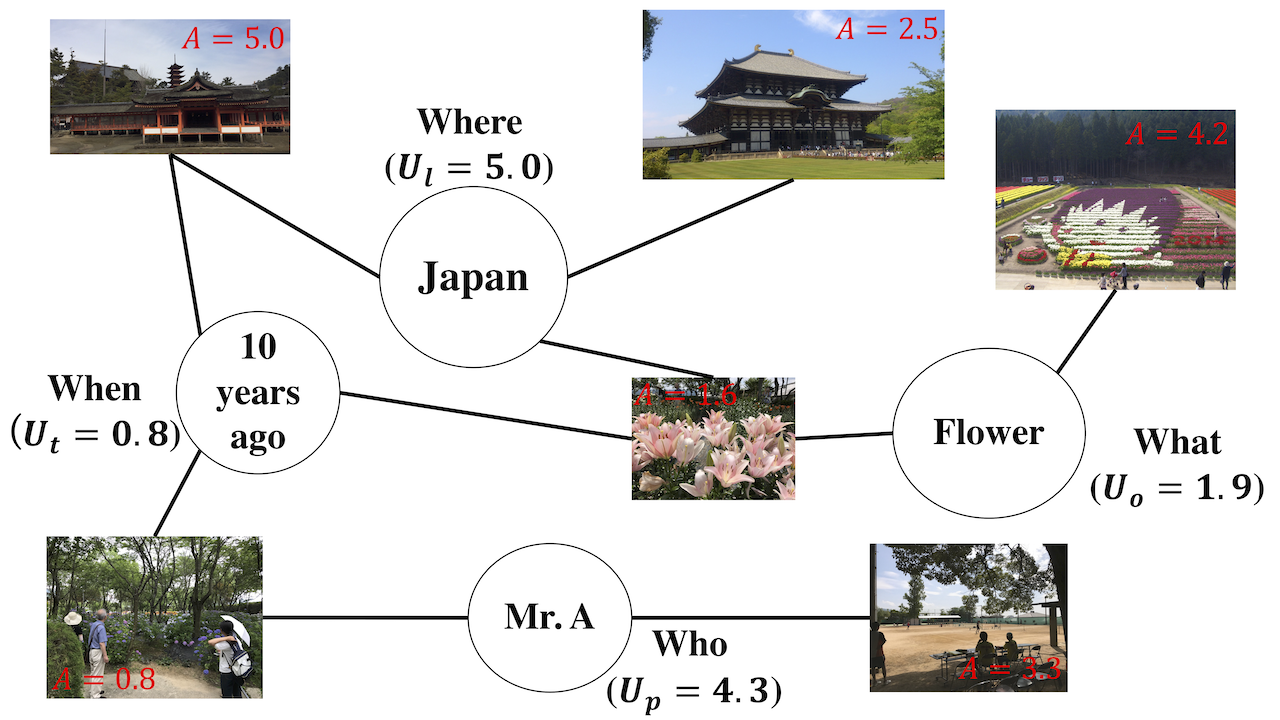}
 \caption{Example of photos network. The number \textit{A} given to each photo means the activation. The photos are connected to each other via a network of attributes. Each attribute has a utility $U_i$ that is assigned to the corresponding production rule.}
 \label{fig:network}
 \vspace{-2mm}
\end{figure}

This framework is applied to the model-based reminiscence described in the previous section: the stimulus ($S$) corresponds to the display photo, and the human response ($R$) corresponds to verbal/nonverbal reactions to the photo.
Although a human internal state ($I$) cannot be directly observed, the self-rating of the mood state by the user can be obtained.
In contrast, the internal state of the model ($I'$) is considered a parameter that can be observed by the modeler.
The parameters consisted in the model-based reminiscence include the activation $A_i$ and the utility $U_i$, which are defined as follows:
%この枠組を，前節で述べたモデルベース回想法に適用した場合，モデルの与える刺激 ($S$) は提示写真，人間の反応 ($R$) は発話に相当する．人間の内部状態 ($I$) は直接観測できないものの，参加者自身の気分の状態の評定が内省報告として得られる．その一方で，モデルの内部状態 ($I'$) はシステム設計者にとって把握可能なパラメータとなる．モデルベース回想法における具体的なパラメータは，活性値$A_i$とユーティリティ$U_i$であり，それぞれ次のように定められる．

%The production rule $i$ for memory retrieval has four attributes: time, place, person, and object in the picture.
%As mentioned in Section \ref{sub2b}, each of the photos in the model's declarative knowledge has the four types of attributes, Who, What, Where, and When.
\begin{itemize}
\item {\it Utility:}
As mentioned in Section \ref{sub2b}, each of the photos in the declarative knowledge of the model consists of  four types of attributes, Who (p: \textbf{p}erson), What (o: \textbf{o}bject), Where (l: \textbf{l}ocation), and When (t: \textbf{t}ime).
Fig.~\ref{fig:network} shows an example of a network created from the attributes attached to photos.
As shown in the image, the photos form a network connected by common attributes, and attributes to be used to move to the next photos are determined by the utility of the rule $i$ corresponding to the attribute.
% In the case that Who is represented by person $p$, What is represented by object $o$, Where is represented by location $l$, and When is represented by time $t$, the utility $i$ for a rule $i$ can be expressed as $i\in\{t, l, p, o\}$.
The utility for the rule $i$ can be expressed as:% $i\in\{p, o, l, t\}$:
%セクション\ref{sub2b}で述べたように，モデルの宣言的知識に含まれる写真はそれぞれ，Who, What, Where, When, の属性をもつ．図\ref{fig:network}に写真に付与される属性から作られるネットワークの例を示す．このように写真同士は共通の属性を介してネットワークを形成している．これらの属性のうち，どの属性が共通している写真を検索するかは，属性に対応するプロダクションルール$i$のユーティリティ値によって決まる．これらの属性をそれぞれ，撮影時期 $t$，撮影場所 $l$，写っている人物 $p$，写っている物 $o$とするとプロダクションルール$i$に対するユーティリティ$i$は$i\in\{t, l, p, o\}$と表すことができる．${U_i}^{n}$は前回のユーティリティ値${U_i}^{n-1}$と報酬値${R_i}^{n}$，学習率$\alpha$から，\ref{equ:u}のように計算される．
\begin{equation}
\label{equ:u}
{U_i}^{n}={U_i}^{n-1}+\alpha[{R_i}^{n}-{U_i}^{n-1}] \mid i\in\{p, o, l, t\}
\end{equation}
where ${U_i}^{n}$ is calculated from the previous utility ${U_i}^{n-1}$, the learning rate $\alpha$, and the reward ${R_i}^{n}$ provided by the self-rating of mood by the user. 

\if0
The activation $A_i$ is calculated as the sum of the base-level activation $B_i$, the strength of association $Sa_i$ in the spreading activation, and the noise $\epsilon_i$ as:
%活性値$A_i$はベースレベル活性値$B_i$，活性化拡散$S_i$，ノイズ$\epsilon_i$の合計値として\ref{equ:a}のように計算される．
\begin{equation}
\label{equ:a}
{A_i}={B_i}+{Sa_i}+{\epsilon_i}
\end{equation}
\fi

\item {\it Activation:} Based on the original ACT-R setting \cite{Anderson:2007}, the activation $A_i$ in this study is calculated as:
%活性値$A_i$はベースレベル活性値$B_i$，活性化拡散$S_i$，ノイズ$\epsilon_i$の合計値として\ref{equ:a}のように計算される．
\begin{equation}
\label{equ:a}
{A_i}=\ln(\sum^{n}_{j=1}{t^{-d}_{j}})+{Sa_i}+{\epsilon_i}
\end{equation}
where $n$, $t_j$, and $d$ are the frequency of the recollection of chunk $i$, the time passed since the $j$th recall, and the decay rate, respectively. These parameters induce the concentration on a specific photo \cite{lebiere2009balancing}, whereas $Sa_i$ indicates a term representing the effects of contexts (see \cite{morita2016modelbased} for details). $\epsilon_i$ is a transient noise parameter that mitigates the concentration on a specific photo, which was mapped to the HRV of the user in the previous studies \cite{itabashi2020interactive,morita2022regulating}.
\end{itemize}

\begin{comment}
In this study, we calculate the following base-level activation: 
%本研究では，ベースレベル活性値の計算に式\ref{equ:ba}を用いる．
\begin{equation}
\label{equ:ba}
{B_i}=\ln(\sum^{n}_{j=1}{t^{-d}_{j}})+\beta_i
\end{equation}
where $n$, $t_j$, $d$, and $\beta_i$ denote the frequency of recall of chunk $i$, the time passed since the $j$th recall, the decay rate, and the offset, respectively.
%$n$はチャンク$i$の出現回数，$t_j$は$j$番目の出現からの経過時間，$d$は減衰率，$\beta_i$はオフセット値を示す．

The spreading activation which is the second term in Eq. \ref{equ:a}, is calculated as the strength of association $S_i$ of chunk $i$ with respect to the context $C$ of the current stimulus $S$:
%式\ref{equ:a}の第二項である活性化拡散は現在の刺激$S$のコンテクスト$C$に対するチャンク $i$の連想強度$S_i$として計算される．

\begin{equation}
\label{equ:s}
{Sa_i}=\sum_{j \epsilon C}{W_{i}Sa_{ji}}
\end{equation}
where $C$ denotes the set of attributes $j$ of the current stimulus $S$.
$W_j$ refers to the weight of attention given to attribute value $j$, and $Sa_{ji}$ denotes the association strength between attribute value $j$ and chunk $i$, which are related to the declarative memory.
%$C$は，現在の刺激$S$の属性値$j$の集合を表す．$W_j$は属性値$j$に付与される注意の重みを表し，$S_{ji}$は属性値$j$と宣言的記憶のチャンク$i$の連想強度を表す．
\end{comment}

In this study, we assumed that the stimulus ($S$) provided by the model is determined according to the internal state of the model ($I'$), which indirectly affects the human internal state ($I$).
Based on this relationship, this study assumed that the estimation of the internal state of the model ($I'$) using the human response ($R$) is a means of estimating the human internal state ($I$).
%これらのモデルの内部状態$I'$に応じて，実験参加者に提示される刺激$S$が定まり，間接的に人間の状態$I$に影響が及ぶ．本研究はそのような関係から，人間から観測される反応$R$による$I'$の推定が，人間の内部状態$I$の推定の代替となると考える．
Thus, this study proposed a method that interactively estimates the human internal states based on the model's internal states and human responses to stimuli.
To verify the practicability of this proposed method, we investigated the possibility of estimating the two internal states (i.e., the human and model internal states) from the human response  ($R$).
To achieve this, we set two research questions (RQs):
%．本研究で提案したモデルが人間に与える刺激に対する反応を基に対話的な推定を行うことの実現性を確認するため，人間の反応$R$から人間とモデルの2つの内部状態の推定がそれぞれ可能かどうかを調査する．そのためのリサーチクエスチョンとして以下の２点を設定し，これらを先行研究において得られた発話データから検討する
\begin{enumerate}
    \item[RQ 1] ($R\rightarrow I'$) Is it possible to estimate the internal state of the model from human responses to stimuli?
    \item[RQ 2] ($R\rightarrow I$) Is it possible to estimate the human internal state (obtained by self-reports) from human responses?
\end{enumerate}

These questions were examined based on the verbal utterances as $R$ obtained during the experiment where the ACT-R memory model presented lifelog photos ($S$) to participants.
%刺激をうけた人間の反応からモデルの内部状態を推定すること ($R\rightarrow I'$) は可能かどうか
%人間の反応から人間の（内省報告により得られた）内部状態を推定すること ($R\rightarrow I$) は可能かどうか

%%%%%%%%%%%%%%%%%%%%%%%%%%%%%%%%%%%%%%%%%%%%%%%%%%%%%%%%%%%%
\section{Method}
\label{s4}
%\section{方法}

This section presents the investigation method for the feasibility of the proposed estimation method.
First, the experiment using the model-based reminiscence system is described.
Next, the dataset constructed from the experiment is presented. Lastly, the analytical methods are explained.
%このセクションでは，前節で提案した推定手法の実現性を検討するに行った分析について述べる．はじめにモデルベース回想法のシステムを用いて行った実験とその結果について説明する．セクション4bでは実験データから作成したデータセットと分析に使用した特徴量について紹介する．セクション4cでは分析の方法について説明する．

%%%%%%%%%%%%%%%%%%%%%%%%%%%%%%%%%%%%%%%%%%%%%%%%%%%%%%%%%%%%
\subsection{Experiment}
\label{sub4a}
%\subsection{実験}
%We conducted an online experiment to collect data during the use of the model-based reminiscence described in Section \ref{sub2b}.
%In the experiment, two parameters of the model-based reminiscence were manipulated to compare participants' behavior under four conditions: 2 (no activation calculated vs. activation calculated) x 2 (no reward value calculated vs. reward value calculated) in order to examine changes in the internal state of humans due to differences in the internal state of the model.
To examine the effect of the different internal states of the model on the human internal states, we compared the behavior of participants under four conditions (within participants): 1) no activation and no reward, 2) no activation with reward, 3) activation with no reward, 4) activation with reward.
%\ref{sub2b}で説明したモデルベース回想法を評価するオンライン実験を実施した．実験ではモデルの内部状態による人間の内部状態の変化を検討するために，表\ref{tab:condition}に示す二つのパラメータを操作した，2（活性値算出無 vs.活性値算出有）× 2（報酬値無 vs. 報酬値有）の4条件下で参加者の行動を比較した．

As described in Section \ref{sub2a}, the activation of ACT-R is affected by the frequency of memory recall and by memory decay.
Therefore, under the condition in which the activation was calculated, intensive retrieval of recent and frequently retrieved photos is expected to occur owing to the characteristics of the memories simulated by ACT-R. 
In contrast, when the activation was not calculated, not only recent but also older photos are retrieved.
%セクション\ref{sub2a}で述べたように，ACT-Rの活性値は記憶の利用頻度や記憶の減衰の影響を受ける．そのため，ACT-Rでシミュレーションされる記憶の特性によって一部の写真への集中的な検索が起きると考えられる．反対に，活性値を算出しない条件では，直近だけでなく古い写真への検索も行われる．
In addition, as described in Section \ref{sub2a} and \ref{sub2b}, the utility value determines which attributes are used to retrieve the photos.
Therefore, the condition with the reward induces the retrieval of photos that reflect the mood towards the currently displayed photo.
%セクション\ref{sub2a}や\ref{sub2b}で述べたように，どの属性から写真を検索するかはユーティリティ値によって定まる．そのため，報酬値を算出する条件では，ユーザの現在表示されている写真に対する気分が反映された写真の検索が行われる．
From the combination of the effects of these parameters, the model can be assumed to exhibit the behavior listed in Table  \ref{tab:condition}, where each factor corresponds to the two axes of core affect \cite{russell2003core}.
%これらのパラメータによるモデルの挙動の組み合わせから，各条件でモデルは表 \ref{tab:condition}に示すような動作をすると考えられる．
%For each condition, the model is expected to have the behavior shown in Table \ref{tab:condition}.
%各条件でモデルは表に示すような動作をすると考えられる．

\if0
% Please add the following required packages to your document preamble:
% \usepackage{multirow}
\begin{table*}[htbp]
\caption{Experimental Conditions}
\label{tab:condition}
\centering
\begin{tabular}{|lc|ll|}
\hline
\multicolumn{2}{|l|}{\multirow{}} & \multicolumn{2}{c|}{\textbf{Activation}} \\ \cline{3-4} 
\multicolumn{2}{|l|}{} & \multicolumn{1}{c|}{\textbf{Not calculated}} & \multicolumn{1}{c|}{\textbf{Calculated}} \\ \hline
\multicolumn{1}{|c|}{\multirow{\textbf{Reward}}} & \textbf{Not calculated} & \multicolumn{1}{l|}{\begin{tabular}[c]{@{}l@{}} Search a wide variety of photos equally. \\ Photo transitions that do not respond to participants' moods\end{tabular}} & \begin{tabular}[c]{@{}l@{}}Search biased toward some photos. \\ Photo transitions that do not respond to participants' moods\end{tabular} \\ \cline{2-4} 
\multicolumn{1}{|c|}{} & \textbf{Calculated} & \multicolumn{1}{l|}{\begin{tabular}[c]{@{}l@{}}Search a wide variety of photos equally.\\ Photo transitions that respond to participants' moods.\end{tabular}} & \begin{tabular}[c]{@{}l@{}}Search biased toward some photos.\\ Photo transitions that respond to participants' moods.\end{tabular} \\ \hline
\end{tabular}
\end{table*}
\fi

\begin{table*}[htbp]
\caption{Experimental Conditions}
\label{tab:condition}
\centering
\begin{tabular}{|lc|ll|}
\hline
\multicolumn{2}{|l|}{\multirow{1}{*}} & \multicolumn{2}{c|}{Reward (Valence)} \\ \cline{3-4} 
\multicolumn{2}{|l|}{} & \multicolumn{1}{c|}{\textbf{Not calculated}} & \multicolumn{1}{c|}{\textbf{Calculated}} \\ \hline
\multicolumn{1}{|c|}{\multirow{1}{*}{\textbf{Activation }}} & 
\textbf{Calculated} &
\multicolumn{1}{l|}{\begin{tabular}[c]{@{}l@{}} Biased search toward some photos.\\ \relax
Not react to participants' moods\end{tabular}} & \begin{tabular}[c]{@{}l@{}}Biased search toward some photos. \\ \relax
React to participants' moods\end{tabular} \\ \cline{2-4} 
\multicolumn{1}{|c|}{\textbf{(Arousal)}} & \textbf{Not calculated} & \multicolumn{1}{l|}{\begin{tabular}[c]{@{}l@{}}Wide search reaching to a variety of photos.\\ \relax
Not react to participants' moods.\end{tabular}} & \begin{tabular}[c]{@{}l@{}}Wide search reaching to a variety of photos.\\  \relax
React to participants' moods.\end{tabular} \\ \hline
\end{tabular}
\end{table*}

\begin{figure}[tb]
 \centering
 \includegraphics[keepaspectratio, width=50mm]
      {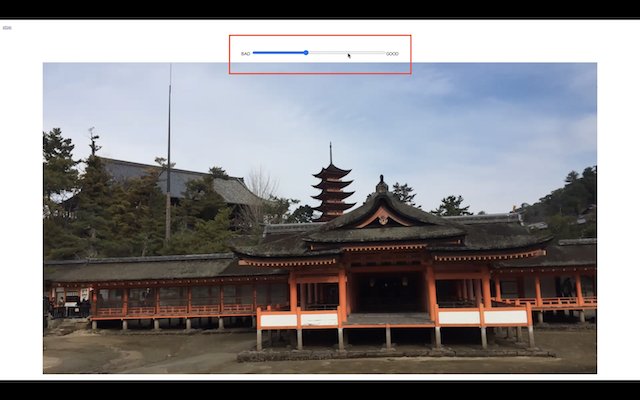}
 \caption{Interface for mood state rating. Users rate their mood state using the interface in the red square.}
 \label{fig:interface}
\end{figure}

For the experiment, twenty-four participants aged \numrange[range-phrase={--}]{21}{61} consisting of \num{13} males and \num{11} females were recruited through the crowdsourcing service Lancers\footnote{https://www.lancers.jp}, and the experiment was conducted using the video chat tool Zoom\footnote{https://zoom.us/}.
%実験参加者は\numrange[range-phrase={ -- }]{21}{61}歳の24名で，男性\num{12}名，女性\num{11}名であった．これらの参加者は，クラウドソーシングサイトランサーズ\footnote{https://www.lancers.jp}にて募集された．実験はビデオチャットツールZoom\footnote{https://zoom.us/}で実施された．
Before participating in the experiment, they provided their personal photos, whose minimum number of photos required to participate in the experiment was 200.
%実験に参加する前に、自分のライフログ写真を提供した。実験に参加するために必要な写真の枚数は最低200枚であった。

%The order for viewing the slideshows constructed with the participants' lifelog in each condition was counterbalanced (consisting of 24 patterns), and the viewing duration was 5 minutes for each condition.
%各条件における参加者のライフログで構成されたスライドショーの視聴順序はカウンターバランス（24パターンで構成）、視聴時間は各条件とも5分であった。

Participants watched all four conditions of the slideshow in the experiment.
As there were four experimental conditions, 24 possible patterns of viewing orders existed.
The 24 participants were assigned to each pattern in order to counterbalance the viewing order of the experimental conditions.
The viewing duration for each condition was 5 min.
%参加者は実験において4つの条件によるスライドショーをすべて閲覧した．
%実験条件は4条件あるため，その順番は24パターン考えられる．
%実験条件の閲覧順によるカウンターバランスを取るため，24名の参加者はそれぞれのパターンに割り当てられた．
%各条件の閲覧時間は5分間であった．

To evaluate the changes in the mood state caused by viewing the photo slideshow, a mood state survey was conducted using the Japanese version of the Profile of Mood States 2nd Edition (POMS-2) \cite{yokoyama1990production} before and after viewing each photo slideshow condition.
The participants were also asked to verbalize their thoughts while viewing the photos \cite{ericsson1980verbal} and rate their current mood on a 6-point scale in response to the photos displayed on the user interface, as indicated by the red boxed position in Fig.~\ref{fig:interface}.
% 写真スライドショーの閲覧による気分状態の変化を評価するために，Profile of Mood States 2nd Edition (POMS-2) の日本語版\cite{yokoyama1990production}による気分状態の調査を写真閲覧前，各条件の閲覧後に実施した．また，参加者は写真閲覧中に思考の発話をすることと，図\ref{fig:interface}の赤枠で囲んだ位置に表示されるユーザインタフェースを使用して提示された写真に対する現在の気分を\numlist[list-final-separator = {，}]{0;2;4;6;8;10}の6段階で評定するタスクが付与された．また，このユーザインタフェースを以降はスライドバーと称する．

%The results showed that retrieval of photos based on the activation led to significant changes in mood ratings.
%However, in their experiment, the parameters that change the behavior of the model were fixed for the activation, and the effect of the reward value was not fully evaluated.
%In order to support an individual's recollection at the right time, it is necessary to set the appropriate parameters in real-time while inferring the individual's internal state from his or her responses to stimuli.
%結果，活性値に基づく写真の検索が，有意な気分評定の変化を導くことを確かめた．しかし，板橋においてはモデルの挙動を変化させるパラメータのうち，活性値については固定され，報酬値についてもその効果を十分に検討できていなかった．個人の回想を適切なタイミングで支援するためには，個人の反応から個人の内部状態を推測しつつ，リアルタイムに適切なパラメータに設定していく必要がある．

\subsection{Data Set}
\label{sub4b}
%\subsection{データセット}
Both the text and voice data of the participants' utterances were used for the analysis.
The text data was constructed using the software tool Vrew\footnote{https://vrew.voyagerx.com/ja/}, which enables the automatic transcription of recorded audio through voice recognition.
After the automatic transcription, typographical errors were corrected manually, and tags indicating the start and end of the experiment and the time the photos were switched were inserted.

Referring to the photo switching time tag, the recorded audio data was segmented.
Following this first segmentation, segments longer than 11 s were further divided into collections of 11 s segments (11 s is the approximate unit time to switch the display photos). The ACT-R model used in this experiment is programmed to retrieve a photo every 11 s. If the model failed in retrieving an image or retrieved the same photo as the previous trial, the displayed photos were not switched. 
%前節の実験の録画データを基に分析で使用するデータを作成した．分析には発話のテキストデータと音声データを用いた．テキストデータは録画データを音声認識による自動文字起こしが可能なツールVrew\footnote{https://vrew.voyagerx.com/ja/}を用いて作成した．自動文字起こしを行った後，誤字の修正，実験の開始・終了及び写真の切り替わりのタイミングを示すタグの挿入を手作業によって行った．音声データは録画データから音声部分を抽出することで作成した．分析のため音声データは提示写真が切り替わる時間の平均値である11秒で分割した．

To estimate the internal states of the model and participants, we extracted prosodic features from the divided audio files using openSMILE, which is an open-source toolkit that can automatically extract features from audio signals and is primarily used for automatic emotion recognition \cite{eyben2010opensmile}.
%openSMILE is open-source toolkit that can automatically extract features from audio signals and is mainly used in the area of automatic emotion recognition.
%分割された音声ファイルにopenSMILEを適用することで韻律特徴を抽出した\cite{eyben2010opensmile}．openSMILEは音声信号から特徴量を自動抽出することができるオープンソースのツールキットで，主に自動感情認識の分野で活用されている．
Particularly, we utilized a feature vector known as the eGeMAPS feature set, which consists of a total of 88 features: 62 minimalistic parameter sets and 26 extended parameter sets \cite{eyben2015geneva}.
Subsequently, a score obtained from the sentiments analysis using Google's Cloud Natural Language API\footnote{https://cloud.google.com/natural-language} applied to the transcribed texts was added to the features and an 89-dimensional feature vector was constructed for the analysis.

The total number of utterance feature vectors created by the aforementioned procedure was 2403 (an average of 100.125 for the 24 participants). These feature vectors were standardized before analysis.
%韻律特徴の分析では共通の特徴ベクトルを用いる．使用する特徴量はopenSMILEで提供されている特徴量セットのうち，eGeMAPSを用いるものである\cite{eyben2015geneva}．eGeMAPSでは，62種類のミニマリスティックパラメータセットと26種類の拡張パラメータセットの合計88種類の特徴量が提供される．これらに，音声に対応する発話内容の感情分析の結果の数値を加えた89次元の特徴ベクトルを分析に用いた．発話内容の感情分析にはGoogleのCloud Natural Language API\footnote{https://cloud.google.com/natural language}の感情分析機能を用いた．また，特徴ベクトルは事前に標準化を行った．上記の手順で作成した発話の特徴ベクトルの総数は2402であった（参加者24名の平均100.125）．

%%%%%%%%%%%%%%%%%%%%%%%%%%%%%%%%%%%%%%%%%%%%%%%%%%%%%%%%%%%%
\subsection{Analysis Method}
\label{sub4c}
%\subsection{分析手法}
%We analyzed the data based on the two research questions.
Using the as-obtained feature vectors, the internal states of both the model and the participants were classified, and RQs described in Section \ref{s3}  were addressed using the results as follows:
%発話から機械的に得られる特徴から2つのリサーチクエスチョンについての可能性を検討する．そのために音声データから特徴ベクトルを構成し，SVM を用いてモデルの内部状態の分類を行った．

\begin{itemize}
\item {\it Analysis for RQ 1}: The experimental conditions were classified according to the utterances of the participants.
%Since the experiment condition manipulated the activation and the reward as the parameters of the model's internal state, we assume that RQ 1 is available if this classification is possible.
As the parameters of the internal state of the model were manipulated as the experimental condition, which involves the activation and the reward, we assumed that RQ 1 answered ``yes'' if this classification is successful.
%はじめに，リサーチクエスチョン1であるモデルの内部状態の推定のために発話からの実験条件の分類を行った．実験条件ではモデルの内部状態のパラメータに相当する活性値と報酬値を操作したため，この分類が可能であればリサーチクエスチョン1が可能であると考える．

\item {\it Analysis for RQ 2}: Compared to that of the model, the human internal state cannot be directly observed.
Therefore, we classified the data obtained by indirect means (i.e., the self-reports and the self-rating of their moods using the user interface). This analysis was composed of two types of analyses: the first analysis is on the change in the mood ratings measured after the end of each condition, and the other is on the change in the mood ratings for each photo assigned as a task during the photo viewing.
The latter analysis is expected to better reflect the real-time changes in the mood of participants.
\end{itemize}
% In this analysis, we classified mood changes according to the experimental condition, and the mood changes for each photo were more similar to the participants' real-time changes in mood.

%次にリサーチクエスチョン2である人間の内部状態の推定のための分析を行った．人間の内部状態は直接観測することはできない．そのため，間接的な測定に当たる内省法による測定やユーザインタフェースによる自身による自己 の気分の評定値を分類した． この分析では実験条件ごとの変化の分類と，写真一枚ごとのよりリアルタイムに近い気分の変化の分類を行った．

For this analysis, a support vector machine (SVM) was used with the parameters optimized by the grid search. The kernel and parameter ranges are shown in Table \ref{tab:gsparam}.
The evaluation was based on the accuracy and F-measure of the 5-fold cross-validation.
% The accuracy and the F-measure were calculated from the available classification results shown in Table \ref{tab:result} and Eq. (\ref{equ:acc}) -- (\ref{equ:f}).
According to the classification results shown in Table \ref{tab:result}, the accuracy was calculated using: $\frac{ TP + TN }{ TP + FP + FN + TN }$.
The F-measure was calculated $\frac{ 2 \cdot Precision \cdot Recall }{ Precision + Recall }$ using Precision ($\frac{ TP }{ TP + FP }$) and Recall ($\frac{ TP }{ TP + FN }$).
%分析の際，SVMのパラメータの最適化のため，表\ref{tab:gsparam}に示すカーネルとパラメータの範囲でグリッドサーチを行った．評価は 5-fold のクロスバリデーションを行った際の正解率 (Accuracy) とF値 (F-measure) を用いた．正解率とF値については表 \ref{tab:result}に示す取りうる分類結果と式\ref{equ:acc}~\ref{equ:f}から求める．
% 表\ref{tab:result}に示すような分類結果から，正解率は$Accuracy = \frac{ TP + TN }{ TP + FP + FN + TN }$のように計算した．
%F値はPrecision (= \frac{ TP }{ TP + FP })とRecall (= \frac{ TP }{ TP + FN })を用いてF-measure = \frac{ 2 \cdot Precision \cdot Recall }{ Precision + Recall }のように計算した．

\begin{table}[tb]
  \caption{Kernels and Parameters Used During the Grid Search.}
  \label{tab:gsparam}
  \centering
  \scalebox{0.95}{
  \begin{tabular}{lcc}
    \hline
    \textbf{Kernel}  &  \textbf{C}   &  \textbf{Gamma}   \\
    \hline \hline
    Linear  &  \numlist[list-final-separator = {, }]{0.001; 0.01; 0.1; 1; 10; 100}  & -- \\
    RBF  &  \numlist[list-final-separator = {, }]{0.001; 0.01; 0.1; 1; 10; 100}  &  \numlist[list-final-separator = {, }]{0.0001; 0.001; 0.01; 0.1; 1; 10} \\
    \hline
  \end{tabular}
  }
\end{table}

\begin{table}[tb]
  \caption{Available Classification Result}
  \label{tab:result}
  \centering
        \begin{tabular}{|lc|cc|}
        \hline
        \multicolumn{2}{|l|}{\multirow{1}{*}}  &  \multicolumn{2}{c|}{\textbf{True Value}}  \\ \cline{3-4} 
        \multicolumn{2}{|l|}{}  &  \multicolumn{1}{c|}{\textbf{True}}  &  \textbf{False}  \\ \hline
        \multicolumn{1}{|c|}{\multirow{1}{*}{\textbf{Classification Result}}}  &  \textbf{True}  & \multicolumn{1}{c|}{\begin{tabular}[c]{@{}c@{}}TP  \\ (True Positive)\end{tabular}}  & \begin{tabular}[c]{@{}c@{}}FP  \\ (False Positive)\end{tabular}  \\ \cline{2-4} 
        \multicolumn{1}{|c|}{}  & \textbf{False} & \multicolumn{1}{c|}{\begin{tabular}[c]{@{}c@{}}FN  \\ (False Negative)\end{tabular}} & \begin{tabular}[c]{@{}c@{}}TF  \\ (True Negative)\end{tabular}  \\ \hline
    \end{tabular}
\end{table}

%\begin{equation}
%\label{equ:acc}
%Accuracy = \frac{ TP + TN }{ TP + FP + FN + TN }
%\end{equation}

%\begin{equation}
%\label{equ:pre}
%Precision = \frac{ TP }{ TP + FP }
%\end{equation}

%\begin{equation}
%\label{equ:rec}
%Recall = \frac{ TP }{ TP + FN }
%\end{equation}

%\begin{equation}
%\label{equ:f}
%F-measure = \frac{ 2 \cdot Precision \cdot Recall }{ Precision + Recall }
%\end{equation}

%%%%%%%%%%%%%%%%%%%%%%%%%%%%%%%%%%%%%%%%%%%%%%%%%%%%%%%%%%%%
\section{Result}
\label{s5}
%\section{分析結果}
This section presents the results of our analysis regarding the two RQs.
Sections \ref{sub5a} and \ref{sub5b} present the results of the analyses of RQ 1 (i.e., estimating the model's internal state from human responses) and RQ 2 (i.e., estimating the human internal state from his / her reactions), respectively. Before presenting these analyses, Section \ref{resultsE} presents the basic data on the model and human behavior obtained in the experiment.
%個々では前節で述べた2つのリサーチクエスチョンに対する分析の結果を示す．はじめにセクション \ref{sub5a}でリサーチクエスチョン1に対する結果を示す．次にセクション \subsection{5b}でリサーチクエスチョン2に対する結果を示す．

%%%%%%%%%%%%%%%%%%%%%%%%%%%%%%%%%%%%%%%%%%%%%%%%%%%%%%%%%%%%
\subsection{Basic Results of the Experiment}
\label{resultsE}
%\subsection{実験結果}
\subsubsection{Model behavior}
%\subsubsection{モデルの振る舞い}

\begin{figure}[tb]
\centering
    \begin{tabular}{c}
        \begin{minipage}[b]{0.45\linewidth}
            \includegraphics[keepaspectratio, scale=0.25]{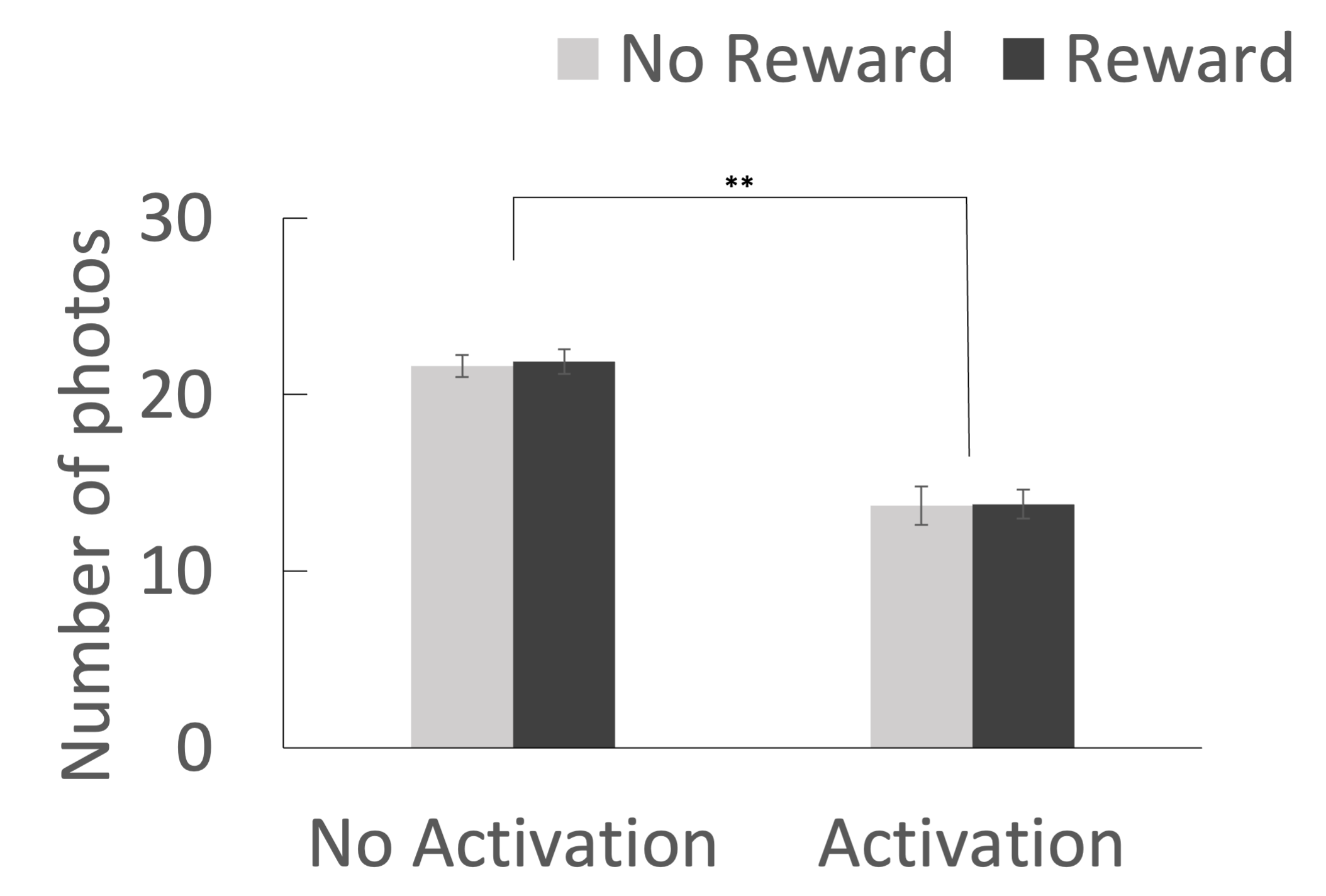}
            \caption{Number of photos presented (Error bars are standard deviation).}
            \label{fig:num_image}
        \end{minipage}
        \hspace{0.01\columnwidth} 
        \begin{minipage}[b]{0.45\linewidth}
            \includegraphics[keepaspectratio, scale=0.25]{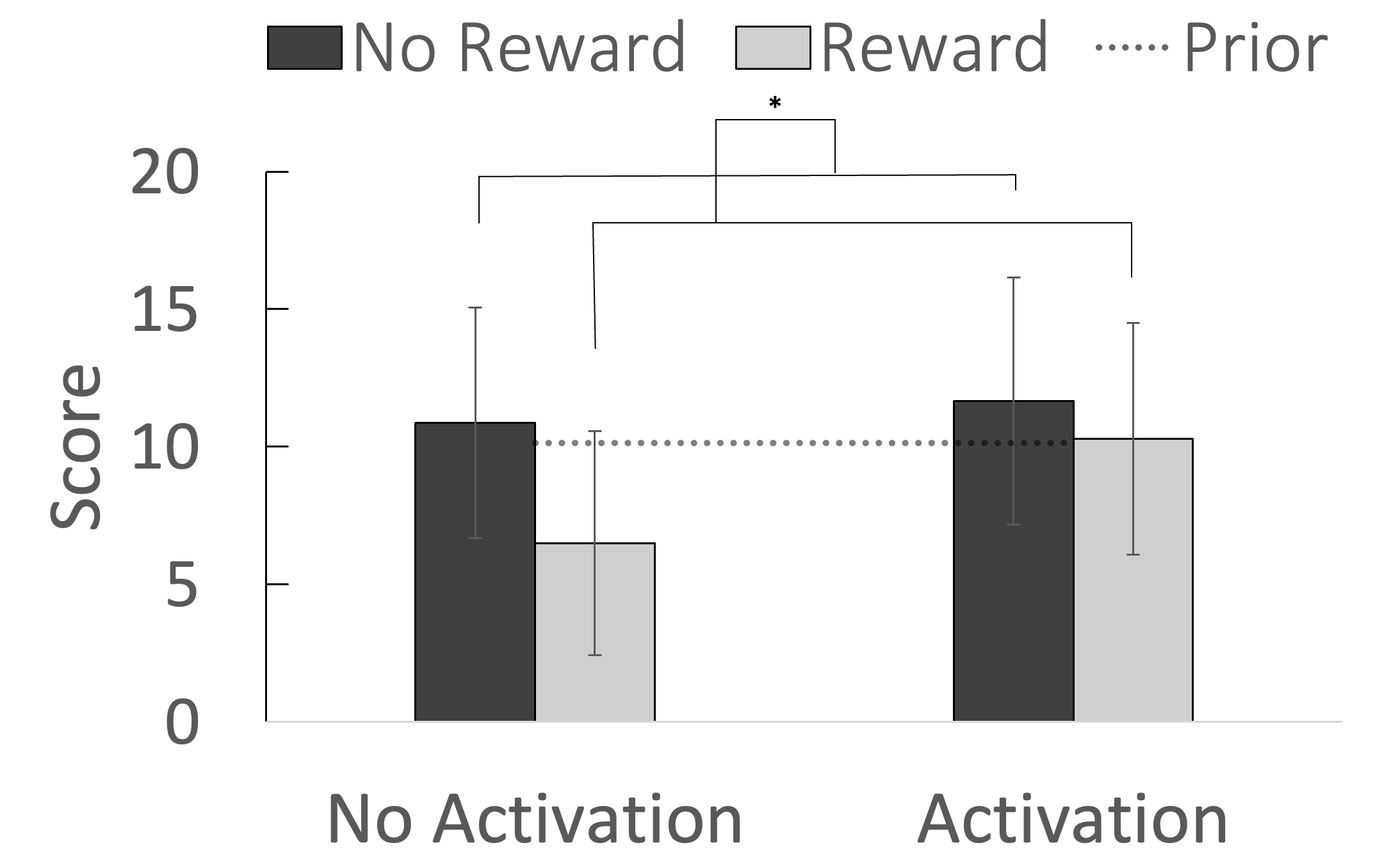}
            \caption{TMD scores for POMS-2 (error bars are standard error).}
            \label{fig:poms_result}
        \end{minipage}
    \end{tabular}
    % \caption{Classification results for each parameters.}
\end{figure}

Fig.~\ref{fig:num_image} shows the results of the number of photos (types) for each of the experimental conditions.
In this graph, the photos that were displayed more than once were counted as one to eliminate duplicates, implying that a smaller value indicates that the model focused on smaller photo sets.
A two-way (the activation vs. the reward) analysis of variance (ANOVA) was conducted using the number of photos presented as the dependent variable.
The analysis revealed that there was a significant difference in the activation ($F(1, 23)=95.48, p < .01$).
%No significant difference in the reward ($F(1, 23)=0.07, n.s.$) and interaction effects ($F(1, 23)=0.02, n.s.$) was found. 
This result confirmed the effect of activation explained in Section \ref{s3}. The analysis in Section \ref{sub5a} presents the examination of the detectability of such model difference from the utterance during the task.
% 図\ref{fig:num_image}に各提示条件における提示写真の枚数（種類）の結果を示す．ただし，複数回表示された写真については，重複を除くため1カウントとした．各パラメータの有無を要因とし，写真の提示枚数を従属変数とする二元配置分散分析を行った．その分析の結果，活性値の主効果が有意であった[$F(1, 23)=95.48, p < .01$].報酬値と交互作用については有意差は見られなかった．

%\begin{figure}[tb]
% \centering
% \includegraphics[keepaspectratio, %width=60mm]{images/image_num.png}
% \caption{Number of photos presented (Error bars are standard deviation).}
% \label{fig:num_image}
%\end{figure}

\subsubsection{Mood changes of the participants}
%\subsubsection{参加者の気分変化}
Fig.~\ref{fig:poms_result} shows the total mood disturbance (TMD) scores, which is a measure of the overall mood state in POMS-2.
The dashed line in the figure corresponds to the TMD score when measured prior to the start of the experiment.
A two-way (the activation vs.the reward) ANOVA with the TMD scores as the dependent variable indicated a significant main effect in the reward ($F(1, 23)=4.59, p < .05$).
As the TMD score indicates that the negativity of a mood increases with an increase in the score, these results suggest that the participants were induced to feel more positive under the reward calculation condition than under the no reward condition by reflecting the mood of the participants to the model. The possibility of detecting such mood changes from the utterance during the task is explored in Section \ref{sub5b}.
%図 \ref{fig:poms_result} に POMS-2 における総合的気分状態の尺度となる TMD 得点の結果を示す．図の点線は，写真閲覧実験開始前に実施したときのTMD得点である．TMD得点を従属変数とし，活性値と報酬値の有無を要因とする二元配置分散分析を行った．その結果，報酬値の主効果が有意であった ($F(1, 23)=4.59，p < .05$).TMD得点は得点が高いほどネガティブな傾向が強いことを示し指標である．分析の結果から実験参加者は報酬値算出条件では，ユーザの気分をモデルに反映させることで報酬値無し条件と比べて，ポジティブな気分へ誘導されたことが示唆される．

%\begin{figure}[tb]
% \centering
% \includegraphics[keepaspectratio, width=60mm]{images/poms_result.png}
% \caption{TMD scores for POMS-2 (error bars are standard error).}
% \label{fig:poms_result}
%\end{figure}
%%%%%%%%%%%%%%%%%%%%%%%%%%%%%%%%%%%%%%%%%%%%%%%%%%%%%%%%%%%%

\subsection{Classification of the Internal States of the Model}
\label{sub5a}
%\subsection{実験条件の分類}
\subsubsection{Combination of the two parameters}
\begin{figure}[tb]
 \centering
 \includegraphics[keepaspectratio, scale=0.35]{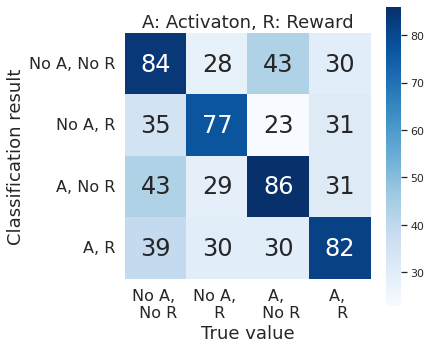}
 \caption{{Classification results of the experimental conditions. (A: Activation, R: Reward)}}
 \label{fig:condition_heatmap}
\end{figure}

To address RQ 1, the four conditions, which involve the presence or absence of activation and rewards, were classified.
%This classification describes the differences in the model's internal parameters, and it corresponds to an estimation of the model's internal state.
%We believe that if this classification is possible, it would provide an answer to RQ1.
%はじめに，（活性値算出無 vs. 活性値算出有）×（報酬値無vs.報酬値有）の4条件の分類を行った．これはモデルの内部パラメータの差を分類するものであり，モデルの内部状態の推定に相当するものである．この分類が可能であれば本研究のRQ1に対する回答が得られると考える．
The result revealed that the classification performance exceeded chance level ($Accuracy_{conditions} = 0.456$, $ F_{conditions} =0.457$).
%実験条件の分類の正解率は0.456，F値は0.457であった．
These results and the heat map shown in Fig.~\ref{fig:condition_heatmap} suggest the possibility of estimating differences in the internal state of the model using utterances from the user.
% この結果と図6に示すヒートマップからわかるように，音声からモデルの内部状態の違いを推定することの可能性が示唆された．
%Although neither the accuracy nor the F-measure was high, the confusion matrix presented in Fig.~\ref{fig:condition_heatmap} shows that for all four true conditions, the corresponding classification results obtained the highest cases. 
%正解率，F値ともに高い値とはならなかったが，混合行列を見るとSVMによる判断の結果の内，4条件全てで真の値に分類されている割合が一番高くなっているため，分類の傾向が見られることがわかる．
%This result suggests the possibility that the model's internal state can be estimated from the user's utterance.
%分類結果の混合行列のヒートマップを図\ref{fig:condition_heatmap}に示す．この結果は、ユーザーの発話からモデルの内部状態を推定できる可能性を示唆しています。

\subsubsection{Independent parameters}
\begin{figure}[tb]
\centering
    \begin{tabular}{c}
        \begin{minipage}[b]{0.45\linewidth}
            \includegraphics[keepaspectratio, scale=0.35]{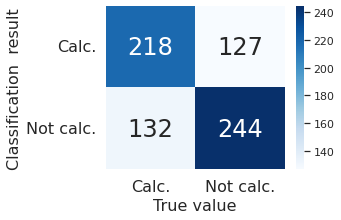}
            \caption{Activation classified result.}
            \label{fig:activation_heatmap}
        \end{minipage}
        \hspace{0.01\columnwidth} 
        \begin{minipage}[b]{0.45\linewidth}
            \includegraphics[keepaspectratio, scale=0.35]{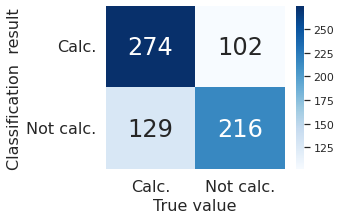}
            \caption{Reward classified result.}
            \label{fig:reward_heatmap}
        \end{minipage}
    \end{tabular}
    % \caption{Classification results for each parameters.}
\end{figure}

Next, we further examined how well each of the internal states of the model (activation and reward) can be classified from the response of the participants. Thus, the analysis consisted of two classifications: classification of the ``calculated'' or ''not calculated'' activation, and classification of the ``calculated'' or ``not calculated'' the reward.
The confusion matrices of these classifications are shown in Figs. \ref{fig:activation_heatmap} and \ref{fig:reward_heatmap}.
%発話からモデルの内部状態を分類する可能性が示唆されたため、活性化と報酬という2つのパラメータをそれぞれ個別に分類した。上記の4条件の内，活性値の有無についての分類と報酬値の有無についての分類をそれぞれ行った．その結果，活性値の分類の正解率は0.634，F値は0.634，報酬値の分類の正解率は0.652，F値は0.652であった．分類結果の混合行列のヒートマップをそれぞれ図\ref{fig:activation_heatmap}，\ref{fig:reward_heatmap}に示す．
Both matrices exhibited moderate classification results ($Accuracy_{A}= 0.640$, $F_A = 0.640$, $Accuracy_R=0.680$, $F_R= 0.679$).
% これらの結果から，発話から写真閲覧時の活性値の高低，報酬値の有無による写真提示の違いに対する反応の違いを検出することの可能性が示唆される．
From these results and the relationships discussed in Section \ref{s3}, the differences in the stimuli provided by the model were considered to affect the human internal state, and those differences in the internal state were reflected in the responses (utterance).
%これらの結果とセクション3で述べた仮定から，モデルの与える刺激の違いが人間の内部状態に影響を与え，その差が反応である発話に現れていると考える．

%%%%%%%%%%%%%%%%%%%%%%%%%%%%%%%%%%%%%%%%%%%%%%%%%%%%%%%%%%%%
\subsection{Classification of Human Internal States}
\label{sub5b}
\subsubsection{Classification based on post-hoc self-report}
%\subsection{POMS-2の結果の変化の分類}

\begin{figure}[tb]
\centering
    \begin{tabular}{c}
        \begin{minipage}[b]{0.45\linewidth}
            \includegraphics[keepaspectratio, scale=0.35]{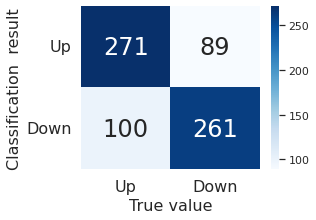}
            \caption{Classification of changes in TMD scores on the POS-2.}
            \label{fig:poms_heatmap}
        \end{minipage}
        \hspace{0.01\columnwidth} 
        \begin{minipage}[b]{0.45\linewidth}
            \includegraphics[keepaspectratio, scale=0.35]{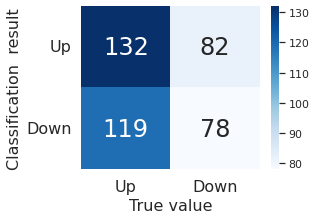}
            \caption{Classification of changes in mood rating.}
            \label{fig:slideupdown_heatmap}
        \end{minipage}
    \end{tabular}
    % \caption{Classification results for each parameters.}
\end{figure}

This analysis investigated the feasibility of estimating the human internal state from their utterances (RQ2).
%The POMS-2 is a commonly used mood scale in which the total mood state is measured as the total mood disturbance score (TMD score).
In this analysis, we classified whether the TMD scores of the POMS-2, which were answered after viewing the photos, were higher or lower than those before viewing or after viewing one of the previous conditions.
However, two participants whose scores kept decreasing during the experiment were excluded from this analysis.
The results are shown in Fig.~\ref{fig:poms_heatmap}, and they indicate that the classification performance for the human internal state was better than that for the internal state of the model ($Accuracy_{TMD}=0.738$, $F_{TMD}=0.738$).
This result suggests the possibility of estimating changes in the mood of users from their utterances.
%この分析ではユーザの発話からユーザの内部状態を推定することの実現性を検討する．ユーザの気分の変化を示す指標としてPOMS-2の結果を利用した．POMS-2は標準的に利用されている気分の尺度であり，総合的な気分状態がTMD (total mood disturbance) 得点として測定される．写真閲覧後に回答した POMS-2 のTMD得点が閲覧前，もしくは一つ前の条件の閲覧後と比べて上がっているか下がっているかを分類した．ただし，2名は4回の閲覧で下がり続けたため除外した．TMD得点の変化の分類の正解率は0.731，F値は0.731であった．分類結果の混合行列のヒートマップを図\ref{fig:poms_heatmap}に示す．活性値の有無，報酬値の有無の分類の場合よりも正解率，F値が共に向上し，混合行列を見てもより良い分類ができていることが示されている．ここから発話から参加者の気分変動の推定が可能なことが示唆される．
 
%\begin{figure}[tb]
%    \centering
%    \includegraphics[keepaspectratio, scale=0.5]{images/poms_heatmap.png}
%    \caption{Classification of changes in TMD scores on the POS-2.}
%    \label{fig:poms_heatmap}
%\end{figure}

%%%%%%%%%%%%%%%%%%%%%%%%%%%%%%%%%%%%%%%%%%%%%%%%%%%%%%%%%%%%
\subsubsection{Classification based on real-time self-report}
\label{sub4i}
%\subsection{スライドバーの値の変化の分類}

Next, we examined the possibility of the real-time estimation of the human internal state.
This analysis classified the changes in the mood rating during the photo presentation.
Each utterance segment was labeled with ``up'' and ``down'' direction tags to indicate whether the participants reported that their feeling moved positively or negatively, respectively. 
Data segments in which there was no change were excluded.
The resulting confusion matrix is shown in Fig.~\ref{fig:slideupdown_heatmap}.
The classification performance was not better than that of chance ($Accuracy_{Mood Rating}=0.511$, $F_{Mood Rating}=0.505$). 
This result may be partly attributed to the fact that some participants made few moves on the user interface during the experiment.
%最後にリアルタイムなユーザの内部状態の推定の可能性を検討する．スライドバーの評定値が一つ前の提示写真の評定値から上がったか下がったかを分類した．ただし，変化のなかった場合を除いている．気分評定の変化の分類の正解率は0.501，F値は0.505であった．分類結果の混合行列のヒートマップを図\ref{fig:slideupdown_heatmap}に示す．分類の結果，上昇下降の分類はできていないことが示された．これは，参加者によってスライドバーをほとんど操作していない場合があったためと考えられる．

\subsection{Classification for Individual Participants}
The previous classification included all utterance segments without distinguishing participants.
Considering the characteristics of prosody and emotion change, it is reasonable to assume that individual differences distorted the classification results.
Therefore, in this section, we presented the classification results for each participant.
Table \ref{tab:list} summarizes the results of the model and human internal states for all participants.
As indicated in the mean scores, although there was no difference in the general patterns of the performance, there were improvements in the classification results in this analysis. 
The classification results for POMS-2 (post-hoc self-reports) were higher than those for the two model parameters (activation and reward).
The classification for the real-time human internal states exhibited the worst classification performance.
%これまでの分類では，参加者個人を区別せずにすべての発話セグメントを使用していた．韻律や感情の変化の特徴を考慮すると，個人差が分類結果に歪みを与えていると考えられる．
%そこで，本節では，参加者ごとに学習を行い分類した結果を示す．表\ref{tab:acclist}は，参加者のモデルと人間の内部状態に対する分類結果をまとめたものである．平均正解率に示されるように，指標間の分類性能の一般的なパターンに違いはないものの，今回の分析で分類結果が改善されていることがわかる．POMS-2に対する分類は，2つのモデルパラメータに対する分類より高い．リアルタイムな人間の内部状態の分類は，最も悪い分類性能である．

% Please add the following required packages to your document preamble:
% \usepackage{multirow}
\begin{table*}[tb]
\caption{Individual Classification Result}
\centering
\scalebox{0.85}{
\begin{tabular}{lcccccccc}
\hline
\multicolumn{1}{c}{\multirow{1}{*}{\textbf{ID}}} & \multicolumn{2}{c}{\textbf{Activation}} & \multicolumn{2}{c}{\textbf{Reward}} & \multicolumn{2}{c}{\textbf{POMS-2}} & \multicolumn{2}{c}{\textbf{Mood Rating}} \\
\multicolumn{1}{c}{} & \multicolumn{1}{r}{\textbf{Accuracy}} & \multicolumn{1}{r}{\textbf{F-measure}} & \multicolumn{1}{r}{\textbf{Accuracy}} & \multicolumn{1}{r}{\textbf{F-measure}} & \multicolumn{1}{r}{\textbf{Accuracy}} & \multicolumn{1}{r}{\textbf{F-measure}} & \multicolumn{1}{r}{\textbf{Accuracy}} & \multicolumn{1}{r}{\textbf{F-measure}} \\
\hline \hline
1 & 0.64 & 0.71 & 0.57 & 0.68 & 0.64 & 0.74 & 0.78 & 0.75 \\
2 & 0.87 & 0.84 & 0.89 & 0.76 & 0.77 & 0.81 & 0.67 & 0.58 \\
3 & 0.89 & 0.87 & 0.76 & 0.74 & 0.67 & 0.73 & 0.49 & 0.52 \\
4 & 0.75 & 0.74 & 0.58 & 0.52 & -- & -- & 0.40 & 0.45 \\
5 & 0.74 & 0.75 & 0.80 & 0.87 & 1.00 & 1.00 & 0.51 & 0.56 \\
6 & 0.68 & 0.61 & 0.66 & 0.61 & 0.63 & 0.57 & 0.44 & 0.33 \\
7 & 0.71 & 0.56 & 0.94 & 0.96 & 0.70 & 0.66 & 0.39 & 0.40 \\
8 & 0.94 & 0.97 & 0.77 & 0.78 & 0.80 & 0.74 & 0.56 & 0.41 \\
9 & 0.88 & 0.86 & 0.72 & 0.70 & 0.75 & 0.76 & 0.57 & 0.46 \\
10 & 0.73 & 0.71 & 0.76 & 0.76 & 0.70 & 0.71 & 0.51 & 0.42 \\
11 & 0.84 & 0.74 & 0.76 & 0.78 & 0.79 & 0.67 & 0.66 & 0.56 \\
12 & 0.67 & 0.59 & 0.70 & 0.69 & 0.74 & 0.63 & 0.86 & 0.73 \\
13 & 0.60 & 0.67 & 0.89 & 0.86 & 0.87 & 0.89 & 0.40 & 0.30 \\
14 & 0.64 & 0.59 & 0.66 & 0.71 & 0.72 & 0.80 & 0.43 & 0.42 \\
15 & 0.80 & 0.80 & 0.74 & 0.74 & 0.59 & 0.59 & 0.43 & 0.43 \\
16 & 0.64 & 0.64 & 0.67 & 0.67 & 0.77 & 0.77 & 0.44 & 0.44 \\
17 & 0.60 & 0.60 & 0.76 & 0.76 & 0.71 & 0.71 & 0.39 & 0.39 \\
18 & 0.63 & 0.63 & 0.74 & 0.74 & 0.79 & 0.79 & 0.46 & 0.46 \\
19 & 0.71 & 0.71 & 0.56 & 0.56 & 0.69 & 0.69 & 0.41 & 0.41 \\
20 & 0.78 & 0.78 & 0.70 & 0.70 & 0.85 & 0.85 & 0.48 & 0.48 \\
21 & 0.72 & 0.72 & 0.75 & 0.75 & 0.96 & 0.96 & 0.56 & 0.56 \\
22 & 0.71 & 0.71 & 0.69 & 0.69 & 0.77 & 0.77 & 0.53 & 0.53 \\
23 & 0.53 & 0.53 & 0.73 & 0.73 & -- & -- & 0.44 & 0.44 \\
24 & 0.75 & 0.75 & 0.60 & 0.60 & 0.77 & 0.77 & 0.65 & 0.65 \\
\hline \hline
Mean & 0.73 & 0.71 & 0.72 & 0.72 & 0.76 & 0.75 & 0.52 & 0.49 \\
\hline
\end{tabular}
\label{tab:list}
}
\vspace{-2mm}
\end{table*}

\section{Conclusion}
%\section{おわりに}

In the first section, we discussed the importance of providing support matched to the human internal state to obtain the appropriate effect of reminiscence therapy.
In this study, we employed a model-based reminiscence with a cognitive architecture to support personalized reminiscence.
To personalize the model, the parameters of the model should be mapped to the human internal state during memory recollection.
Although previous studies have directly mapped physiological indices to corresponding model parameters, the application of these methods under various settings is limited. Therefore, to estimate the human internal state when using model-based reminiscence, we proposed a machine-learning based method, in which the model interactively estimates the human internal state using naturally produced user response (utterance).
%適切な回想法の効果を得るために人間の内部状態に合わせた支援の必要がある．そのことから本研究では，個人に合わせた回想支援のため認知アーキテクチャを用いたモデルベース回想法を利用する．個人に合わせるためには，モデルのパラメータを人間の回想時の内部状態に対応させることが重要である．そこでモデルベース回想法利用時のユーザの発話からユーザの内部状態を推定するために，モデルが人間と対話的に人間の内部状態を推定する手法を提案した．

To verify the feasibility of the proposed method, we classified internal states using the utterances of participants during photo observation.
In this analysis, we attempted to classify the internal state of the model (experimental conditions) and the human internal state (POMS-2, mood rating).
The first classification revealed that although the performance was not high, the internal states of the model can be estimated from human reactions ($R \rightarrow I'$ in Fig.~\ref{fig:image}).
The second analysis revealed that the POMS-2 classification results exhibited better classification than the classification of the internal states of the model.
This result suggests that it is possible to estimate changes in the human mood state from prosodic features, and indicates the possibility of estimating human internal states from the human responses ($R \rightarrow I$ in Fig.~\ref{fig:image}).

%提案手法の実現性を確認するため，モデルベース回想法利用時のユーザの発話を用いた内部状態の分類を行った．この分析では，モデルの内部状態（実験条件），および人間の内部状態（POMS-2，スライドバー）の分類を試みた．前者の結果から，高い性能は得られなかったものの発話からの分類の可能性が示唆された．この結果から図\ref{fig:image}の $R \rightarrow I'$ の推定の可能性が示された．後者のうち，POMS-2の分類の結果は，実験条件の分類よりも性能が向上した．このことから，発話の韻律特徴からユーザの気分状態の変化を推定可能であることが示唆された．この結果は，図\ref{fig:image}の $R \rightarrow I$ の推定の可能性を示すものである．

These results indicate that the two RQs described in this study can be answered.
In addition, these classification results also support the effectiveness of the model-based reminiscence.
Particularly, the fact that we were able to classify differences in the internal state of the model from the human responses indicates that different internal states of the model caused differences in the human responses.
This implies that differences in the internal parameters of the model affected the human internal state. Thus, this study successfully advanced the interactive memory support approach based on cognitive models. We believe that this approach will contribute to the control of memory recall accompanied with emotions.
%加えて，これらの分類結果は，モデルベース回想法の有効性を支持するものである．ユーザの反応からモデルの内部状態の違いを分類できたということは，モデルの内部状態が異なるとユーザの反応に差が生まれたということである．つまり，モデルの内部パラメータの差がユーザの内部状態に影響を与えていたと考えられる．

However, some limitations should be noted.
This study did not consider how the estimation of the internal state of humans and models can be utilized to improve the human-model interaction. Particularly, although the fact that the success of this classification reflects the effectiveness of the model parameter in stimulating the human internal state  was considered, the use of the estimation of the internal state of the model was not described. In the future study, a design for implementing the indices in the framework of interactive memory support should be developed. 

Finally, it should also be noted that the classification of the human internal state is based on indirect indicators. Regarding the means of observing human internal states, post-hoc measurements cannot address changes in the internal state depending on the situation, and real-time rating is problematic in terms of user load. Although these limitations are some of our motivations to utilize natural user response (verbalization), as an experimental study, it is better to collect more accurate user responses in real-time. Future studies should collect physiological data to support the result obtained in this study.

%これらの結果から，本研究の２つのリサーチクエスチョンに対し，限定付きではあるものの可能であるとの回答が得られた．ただし，この限定性については十分に注意する必要がある．モデルの内部状態を推定することで，人間の内部状態を推測することが可能であるという前提は，その妥当性自体を改めて検討する必要がある．また，人間の内部状態を目的変数とした分類も，間接的な指標を用いたものであることに注意する必要がある．
%The difficulty of classification with subjective mood ratings might also reflects the problem of such indirect measurement of internal states. 
% 主観的な気分評定を目的変数とした分類における困難さは，このような内部状態の間接的な測定の問題を示している．
%Taken together, this study confirms the feasibility and challenges of having models interactively estimating the human internal state based on human-model interactions.
%これらを総合し，本研究では，人間とモデルのインタラクションからモデルが人間の内部状態を対話的に推定することの実現性と課題を確認したといえる．

%In the future, it will be necessary to develop interactive systems that guide users while dynamically estimating their internal states based on the above findings.
%In the course of such attempts, we believe that means will be developed to overcome the aforementioned limitations.
%今後は，上記の知見を踏まえつつ，内部状態を動的に推定しつつ，ユーザを誘導するインタラクティブシステムを開発していく必要がある．その試みを行う中で，先述した限界を克服する手段が開発されていくと考える．人間の内部状態を観測する手段に関して，事後での計測は，状況に応じた内部状態の変化に対応できず，リアルタイムでの評定はユーザの負荷の点で問題が生じる．これらの限界を克服するために，より簡便かつリアルタイムにユーザの反応を計測する手段の開発が必要である．

%%%%%%%%%%%%%%%%%%%%%%%%%%%%%%%%%%%%%%%%%%%%%%%%%%%%%%%%%%%%
% \begin{comment}
\section*{Ethical Impact Statement}
% The experiment was reviewed and approved by Research Ethics Committee.
The experiment was reviewed and approved by Shizuoka University Research Ethics Committee.
In the experiments of this study, we informed the participants fully about the handling of their personal data in advance, and collected the data after obtaining their free-will consent.
If participants request that all photos uploaded to the server be deleted, they are informed that they can be completely deleted from the server.
% In addition, because learning with cognitive architectures can capture all parameter transitions, the process is clearer than learning with black-box methods such as neural networks.
%人間を対象とした研究は、静岡大学研究倫理委員会の審査と承認を受けた。本研究の実験では事前に参加者に個人データの取り扱いについて十分な説明を行い，参加者の自由意志による同意を得て収集した．サーバにアップロードしたすべての写真について実験参加者から削除の依頼があった場合にはサーバから完全に削除することができることを伝えている．また，認知アーキテクチャを用いた学習はニューラルネットワークなどによる学習と比べ，すべてのパラメータの遷移を捕捉することができるため，その過程がクリアである．
% \end{comment}
%%%%%%%%%%%%%%%%%%%%%%%%%%%%%%%%%%%%%%%%%%%%%%%%%%%%%%%%%%%%

\bibliographystyle{IEEEtran}
\bibliography{ref.bib}

% Generated by IEEEtran.bst, version: 1.14 (2015/08/26)
\begin{thebibliography}{10}
\providecommand{\url}[1]{#1}
\csname url@samestyle\endcsname
\providecommand{\newblock}{\relax}
\providecommand{\bibinfo}[2]{#2}
\providecommand{\BIBentrySTDinterwordspacing}{\spaceskip=0pt\relax}
\providecommand{\BIBentryALTinterwordstretchfactor}{4}
\providecommand{\BIBentryALTinterwordspacing}{\spaceskip=\fontdimen2\font plus
\BIBentryALTinterwordstretchfactor\fontdimen3\font minus
  \fontdimen4\font\relax}
\providecommand{\BIBforeignlanguage}[2]{{%
\expandafter\ifx\csname l@#1\endcsname\relax
\typeout{** WARNING: IEEEtran.bst: No hyphenation pattern has been}%
\typeout{** loaded for the language `#1'. Using the pattern for}%
\typeout{** the default language instead.}%
\else
\language=\csname l@#1\endcsname
\fi
#2}}
\providecommand{\BIBdecl}{\relax}
\BIBdecl

\bibitem{10665-268688}
{International Health Conference}, ``Constitution of the world health
  organization. 1946.'' \emph{Bulletin of the World Health Organization},
  vol.~80, no.~12, pp. 983 -- 984, 2002.

\bibitem{hepper2014pancultural}
E.~G. Hepper, T.~Wildschut, C.~Sedikides, T.~D. Ritchie, Y.-F. Yung, N.~Hansen,
  G.~Abakoumkin, G.~Arikan, S.~Z. Cisek, D.~B. Demassosso \emph{et~al.},
  ``Pancultural nostalgia: prototypical conceptions across cultures.''
  \emph{Emotion}, vol.~14, no.~4, p. 733, 2014.

\bibitem{sedikides2004nostalgia}
C.~Sedikides, T.~Wildschut, and D.~Baden, ``Nostalgia: Conceptual issues and
  existential functions.'' p. 230, 2004.

\bibitem{butler1963life}
R.~N. Butler, ``The life review: An interpretation of reminiscence in the
  aged,'' \emph{Psychiatry}, vol.~26, no.~1, pp. 65--76, 1963.

\bibitem{hallford2021remembering}
D.~J. Hallford, S.~Hardgrove, M.~Sanam, S.~Oliveira, M.~Pilon, and T.~Duran,
  ``Remembering for resilience: Brief cognitive-reminiscence therapy improves
  psychological resources and mental well-being in young adults,''
  \emph{PsyArXiv}, 2021.

\bibitem{bryant2005using}
F.~B. Bryant, C.~M. Smart, and S.~P. King, ``Using the past to enhance the
  present: Boosting happiness through positive reminiscence,'' \emph{Journal of
  Happiness Studies}, vol.~6, no.~3, pp. 227--260, 2005.

\bibitem{akhoondzadeh2014effect}
G.~Akhoondzadeh, S.~Jalalmanesh, and H.~Hojjati, ``Effect of reminiscence on
  cognitive status and memory of the elderly people,'' \emph{Iranian Journal of
  Psychiatry and Behavioral Sciences}, vol.~8, no.~3, p.~75, 2014.

\bibitem{gonzalez2015reminiscence}
J.~Gonzalez, T.~Mayordomo, M.~Torres, A.~Sales, and J.~C. Mel{\'e}ndez,
  ``Reminiscence and dementia: a therapeutic intervention,''
  \emph{International Psychogeriatrics}, vol.~27, no.~10, pp. 1731--1737, 2015.

\bibitem{otake2009coimagination}
M.~Otake, M.~Kato, T.~Takagi, and H.~Asama, ``Coimagination method:
  Communication support system with collected images and its evaluation via
  memory task,'' in \emph{International Conference on Universal Access in
  Human-Computer Interaction}.\hskip 1em plus 0.5em minus 0.4em\relax Springer,
  2009, pp. 403--411.

\bibitem{russell2003core}
J.~A. Russell, ``Core affect and the psychological construction of emotion.''
  \emph{Psychological Review}, vol. 110, no.~1, p. 145, 2003.

\bibitem{bradley1992remembering}
M.~M. Bradley, M.~K. Greenwald, M.~C. Petry, and P.~J. Lang, ``Remembering
  pictures: pleasure and arousal in memory.'' \emph{Journal of Experimental
  Psychology: Learning, Memory, and Cognition}, vol.~18, no.~2, p. 379, 1992.

\bibitem{mathews1998cognitive}
A.~Mathews and B.~Mackintosh, ``A cognitive model of selective processing in
  anxiety,'' \emph{Cognitive Therapy and Research}, vol.~22, no.~6, pp.
  539--560, 1998.

\bibitem{fum2007cognitive}
D.~Fum, F.~Del~Missier, and A.~Stocco, ``The cognitive modeling of human
  behavior: Why a model is (sometimes) better than 10,000 words,''
  \emph{Cognitive Systems Research}, vol.~8, no.~3, pp. 135--142, 2007.

\bibitem{Kotseruba}
I.~Kotseruba and J.~K. Tsotsos, ``A review of 40 years of cognitive
  architecture research: Focus on perception, attention, learning and
  applications,'' \emph{AI Review}, vol.~53, pp. 17--94, 2020.

\bibitem{Anderson:2007}
J.~R. Anderson, \emph{How Can the Human Mind Occur in the Physical
  Universe}.\hskip 1em plus 0.5em minus 0.4em\relax New York: Oxford University
  Press, 2007.

\bibitem{somerscognitive}
S.~Somers, A.~Oltramari, and C.~Lebiere, ``Cognitive twin: A personal assistant
  embedded in a cognitive architecture,'' in \emph{Proceedings of the 18th
  International Conference on Cognitive Modelling}, 2020.

\bibitem{morita2016modelbased}
J.~Morita, T.~Hirayama, K.~Mase, and K.~Yamada, ``Model-based reminiscence:
  Guiding mental time travel by cognitive modeling,'' in \emph{Proceedings of
  the Fourth International Conference on Human Agent Interaction}, ser. HAI
  '16, 2016, p. 341^^e2^^80^^93344.

\bibitem{schacter2007remembering}
D.~L. Schacter, D.~R. Addis, and R.~L. Buckner, ``Remembering the past to
  imagine the future: the prospective brain,'' \emph{Nature Reviews
  Neuroscience}, vol.~8, no.~9, pp. 657--661, 2007.

\bibitem{itabashi2020interactive}
K.~Itabashi, J.~Morita, T.~Hirayama, K.~Mase, and K.~Yamada, ``Interactive
  model-based reminiscence using a cognitive model and physiological indices,''
  in \emph{Proceedings of the 18th International Conference on Cognitive
  Modelling}, no. 2020, 2020.

\bibitem{anderson1991reflections}
J.~R. Anderson and L.~J. Schooler, ``Reflections of the environment in
  memory,'' \emph{Psychological Science}, vol.~2, no.~6, pp. 396--408, 1991.

\bibitem{lebiere2009balancing}
C.~Lebiere and B.~J. Best, ``Balancing long-term reinforcement and short-term
  inhibition,'' in \emph{Proceedings of the 31st Annual Conference of the
  Cognitive Science Society}, 2009, pp. 2378--2383.

\bibitem{dancy2015using}
C.~L. Dancy, F.~E. Ritter, K.~A. Berry, and L.~C. Klein, ``Using a cognitive
  architecture with a physiological substrate to represent effects of a
  psychological stressor on cognition,'' \emph{Computational and Mathematical
  Organization Theory}, vol.~21, no.~1, pp. 90--114, 2015.

\bibitem{morita2022regulating}
J.~Morita, T.~Pitakchokchai, G.~B. Raj, Y.~Yamamoto, H.~Yuhashi, and
  T.~Koguchi, ``Regulating ruminative web browsing based on the counterbalance
  modeling approach,'' \emph{Frontiers in Artificial Intelligence}, vol.~5,
  2022.

\bibitem{yokoyama1990production}
K.~Yokoyama, S.~Araki, N.~Kawakami, and T.~Tkakeshita, ``{Production of the
  Japanese edition of profile of mood states (POMS): assessment of reliability
  and validity},'' \emph{[Nihon koshu eisei zasshi] Japanese journal of public
  health}, vol.~37, no.~11, pp. 913--918, 1990.

\bibitem{ericsson1980verbal}
K.~A. Ericsson and H.~A. Simon, ``Verbal reports as data.'' \emph{Psychological
  Review}, vol.~87, no.~3, p. 215, 1980.

\bibitem{eyben2010opensmile}
F.~Eyben, M.~W{\"o}llmer, and B.~Schuller, ``{OpenSmile: the munich versatile
  and fast open-source audio feature extractor},'' in \emph{Proceedings of the
  18th ACM International Conference on Multimedia}, 2010, pp. 1459--1462.

\bibitem{eyben2015geneva}
F.~Eyben, K.~R. Scherer, B.~W. Schuller, J.~Sundberg, E.~Andr{\'e}, C.~Busso,
  L.~Y. Devillers, J.~Epps, P.~Laukka, S.~S. Narayanan \emph{et~al.}, ``{The
  Geneva minimalistic acoustic parameter set (GeMAPS) for voice research and
  affective computing},'' \emph{IEEE Transactions on Affective Computing},
  vol.~7, no.~2, pp. 190--202, 2015.

\end{thebibliography}

%\end{thebibliography}

\begin{comment}
\vspace{12pt}
\color{red}
IEEE conference templates contain guidance text for composing and formatting conference papers. Please ensure that all template text is removed from your conference paper prior to submission to the conference. Failure to remove the template text from your paper may result in your paper not being published.
\end{comment}

\end{document}